\documentclass[journal]{IEEEtran}
\usepackage{tikz}
\usetikzlibrary{arrows}
\usetikzlibrary{positioning}
\usepackage{caption}
\usepackage{amsmath, amsthm, amssymb, amsfonts}
\usepackage[ruled,vlined]{algorithm2e}
\usepackage{chngcntr}
\DeclareSymbolFont{matha}{OML}{txmi}{m}{it}
\DeclareMathSymbol{\varv}{\mathord}{matha}{118}
\usepackage{amsmath,array,float}
\usepackage[export]{adjustbox}
\usepackage[utf8]{inputenc}
\usepackage{notoccite}
\usepackage{hyperref}
\usepackage{bm}
\usepackage[bottom]{footmisc}
\usepackage[numbers]{natbib}
\usepackage{float}
\usepackage{ragged2e}       % new
\usepackage{booktabs,       % new
            makecell,       % new
            tabularx}       % new
\usepackage{cellspace}
\usepackage{balance}
\usepackage{soul}
\usepackage[bottom]{footmisc}
\usepackage{svg}

\floatstyle{plaintop}
\restylefloat{table}

\usepackage{soul}
\usepackage{xcolor,graphicx}
\iftrue  % color
  \graphicspath{{img-color/}}
\else % grayscale
  \graphicspath{{img-gray/}}
\fi

\newcommand{\parenthnewln}[1]{\right.\\#1&\left.{}}

\title{\LARGE \bf
Performance vs. Spectral Properties For Single-Sideband Continuous Phase Modulation
}

\author{Karim Kassan, Haïfa Farès, D. Christian Glattli and Yves Louët

\thanks{K. Kassan, H. Far\`es and Y. Louet are with IETR - UMR CNRS 6164, CentraleSup\'elec Rennes Campus, avenue de la Boulaie - CS 47601 35576 CESSON-SEVIGNE Cedex, France (e-mail: karim.kassan@ieee.org \{haifa.fares, yves.louet\}@centralesupelec.fr).

D. C. Glattli is with SPEC UMR 3680 CEA-CNRS, Universit\'{e} Paris-Saclay, CEA-Saclay, 91191 Gif-Sur-Yvette, France (e-mail: christian.glattli@cea.fr).}}

\DeclareUnicodeCharacter{2212}{-}

\begin{document}
\twocolumn[
  \begin{@twocolumnfalse}
  
    \begin{center}
        \huge “This work has been submitted to the IEEE for possible publication. Copyright may be transferred without notice, after which this version may no longer be accessible.” 
      \end{center}
     
  \end{@twocolumnfalse}
]
\pagebreak
\maketitle
\thispagestyle{empty}
\pagestyle{empty}
\begin{abstract}
    This study revokes the performance of continuous phase modulation (CPM) able to generate a single-sideband (SSB) spectrum directly. This signal is analyzed in terms of modulation indices, pulse lengths, and pulse widths, all of which affect error probability, bandwidth, SSB property, and receiver complexity. The error probability performance is based on an approximation of the minimum Euclidean distance. A numerical power spectral density calculation for this particular SSB modulation in terms of modulation index is presented. Reasonable tradeoffs in designing modulation schemes have been proposed using multi-objective optimization to ensure sizable improvements in bit error rate (BER), spectral efficiencies, and complexity without losing the property of being a SSB signal. Performance comparisons are made with known CPM schemes, e.g., Gaussian Minimum Shift Keying (GMSK) and Raised Cosine based CPM (RC). 
    
    % The \textit{M-ary} case, for M from 2 to 8, is also included in this paper.
\end{abstract}

\maketitle
\IEEEdisplaynontitleabstractindextext
\IEEEpeerreviewmaketitle
\section{Introduction}
\IEEEPARstart{C}{ontinuous} phase modulation (CPM)~\cite{anderson2013digital} is particularly suitable for long-distance wireless communications due to spectral efficiency and high energy efficiency as a result of its constant signal envelope. Many digital communication systems use the CPM family extensively for these attractive properties, e.g., satellite communications, deep-space, optical fiber, telemetry, etc. Further to these advantages, the authors in~\cite{fares2018quantum} presented a CPM scheme, which has the original feature of directly generating a single-sideband (SSB) spectrum, which in turn, provides a highly compact frequency occupancy. By \emph{directly,} we mean that SSB property results from the original modulation, and not from post-filtering~\cite{fares2017power}. This waveform is hereafter referred to as Single-Sideband Frequency Shift Keying (SSB-FSK). SSB-FSK uses a generic phase derivative pulse with \emph{Lorentzian} shape and a $2\pi$ phase increment. The Lorentzian pulse belongs to specific shapes addressing fundamental quantum physics, particularly the on-demand injection of a single electron in a quantum conductor. Digital transmissions based on CPM are the first immediate application of \emph{classical levitonics} \cite{DC2016}. In \cite{fares2017power}, we presented the principle of the modulation and its mathematical justification. However, we did not provide any result about detection schemes or the interplay between error probability performance, bandwidth, and SSB property. Besides, CPM signals suffer from receiver complexity due to the memory property introduced by the phase continuity, which should also be considered alongside the error probability, bandwidth, and SSB property. In this paper, we suggest several tradeoffs under all these considerations to get the full potential of SSB-FSK signals.  

\subsection{Related Work On Single-Sideband CPM}
SSB-FSK originates from the theoretical proposal of Levitov \emph{et al.}~\cite{levitov1996electron}, who demonstrated how to create a pure single electron without any perturbation in the quantum conductor. It is widely accepted that a quantum conductor is full of many free-moving electrons. When one electron is excited by an elementary pulse, others tend to move to fill the hole created, generating undesirable neutral plasmonic waves. 
To overcome this issue,  Levitov \emph{et al.} predicted that one could inject an electron without causing any disturbance \emph{if and only if} the voltage pulse follows a Lorentzian time shape.  Using the Lorentzian pulse, we can generate a pure single electron state, called a “Leviton”, a new quasi-particle with minimal excitation. The Levitov idea is given further experimental illustration by the authors in~\cite{dubois2013minimal}, and \cite{jullien2014quantum}, who showed that when using a Lorentzian pulse, the electron energy distribution becomes SSB---the left part of the energy distribution disappears. Based on this experimentation, the authors in~\cite{DC2016} proposed a new CPM modulation signal, which directly provides a SSB signal. Then in~\cite{fares2018quantum} and \cite{fares2017new}, we gave the first system model of SSB-FSK as a CPM modulation signal.\\
Additionally, we introduced properties, such as the symbol-by-symbol coherent demodulation scheme, based on SSB-FSK signals' orthogonality. Due to the CPM signals' memory, we developed an average matched filter detector, making decisions according to a given observation window. We then analyzed the power spectral density (PSD) for the binary case, leading us to demonstrate that the PSD of the SSB-FSK signal is unilateral with respect to the carrier frequency $(f_c)$, and almost all of the power is concentrated in the first period of the frequency band. Finally, in~\cite{fares2017power}, we presented an analytical expression for the power spectral density (PSD) for binary and quaternary SSB-FSK.

\subsection{Main Contributions}
In this paper, we report a complete study of the SSB-FSK performance regarding error probability, spectral efficiency, and SSB property depending on the following parameters:
\begin{itemize}
    \item modulation index $h$: even if increasing $h$ consistently increases the bandwidth ($BW$) occupancy, it does not carry a monotonic impact on the error probability performance and the SSB property preservation.
    \item pulse length $L$: expanding $L$ results in a complete response of the Lorentzian filter and will introduce a better $BW$ occupancy and lower side lobes, although the larger $L$, the larger the receiver complexity.
    \item Modulation level size $M$: increment in $M$ produces a gain in error probability performance while degrading the $BW$ occupancy and increasing complexity.
    \item pulse width $w$: it is a new tuning parameter introduced specially for SSB-FSK, which plays an essential role in error probability and $BW$ occupancy performance.
\end{itemize}
   It is worth emphasizing the need for a functional interplay between all these parameters, which lead to opposite effects on the error probability, spectral efficiency, and complexity. Therefore, we need to define certain tradeoffs between these parameters as long as an optimal scheme considering all performance metrics is impossible.

Five key contributions of this paper are:
\begin{enumerate}
    \item The study of the Maximum Likelihood Sequence Detector (MLSD) performance on the SSB-FSK signals, and the effects of all parameters $(h,w, L, M)$ on the error probability performance---mainly the effect of pulse width $w$. Additionally, we highlight some points related to memory performed by the Viterbi algorithm~\cite{forney1973viterbi}. We gave the error probability performance in terms of an approximation of the achievable minimum Euclidean distance.
    \item The study of the effect of all parameters $(h, w, L, M)$ on the power spectral density (PSD): this study is mainly performed for $h\neq1$, and its results have been compared to those obtained from the particular case of $h=1$ (responsible of pure SSB signal and particularly relevant for synchronization purpose)~\cite{fares2018quantum}. 
    % \textcolor{red}{Here we exhibit a better spectrum efficiency, with $h\neq1$ and $L\leq 12$ (lower receiver complexity) for a variety of $w$, with a slight loss in SSB property.}
    \item A method to select the best parameter combinations to obtain the SSB-FSK scheme's optimal performance; this method is based on the \textit{Pareto optimum}, which is a multi-objective optimization method.
    This study is done in two steps: first, the optimization is performed with no constraints on the receiver complexity. Second, this complexity is considered the third objective function, alongside energy efficiency and bandwidth occupancy.
    \item Energy-bandwidth comparisons for SSB-FSK signals, where we were able to define a parameter configuration, with acceptable tradeoffs between energy consumption, $BW$ occupancy, and complexity for SSB-FSK that outperforms popular CPM schemes (e.g., GMSK, RC).
    A new SSB-FSK configuration with an integer modulation index $h$ combines good performance and advantage in synchronization.
\end{enumerate}
The rest of the paper is organized as follows. In Section \ref{sec2}, we briefly introduce the signal model. In Section \ref{sec3}, we evaluate the error probability performance of SSB-FSK signals using the union bound. In Section~\ref{sec4}, we present a numerical method to compute the power spectral density, using the autocorrelation approach. In Section~\ref{sec5}, we present the \textit{Pareto optimum}, a multi-objective optimization method, which is used alongside a \textit{brute force} to obtain the best possible tradeoffs for SSB-FSK scheme. In Section~\ref{sec6}, we illustrate several simulation results for error probability performance, power spectrum performance, energy-bandwidth comparison, and synchronization advantage respectively. Finally,the outcomes of this study in terms of concluding remarks and design directives are summarized in Section \ref{sec+}. Conclusions are drawn in Section \ref{sec7}.

\section{Single-Sideband CPM System Model}
\label{sec2}

\begin{table}[]
    \centering
    \captionsetup{justification=centering}
    \caption{\\Table of symbols.}
    \resizebox{0.9\columnwidth}{!}{%    
    \begin{tabular}{cc}
        \toprule
        Symbol                               & Indication                                                                                                                       \\ \midrule \midrule
        $h$, $\tilde{h}$                     & \begin{tabular}[c]{@{}c@{}}CPM modulation index, \\ SSB-FSK CPM modulation index\end{tabular}                                    \\ \hline
        $L$                                    & Pulse length                                                                                                                     \\ \hline
        $M$                                    & Modulation level                                                                                                                 \\ \hline
        $w$                                    & Pulse width                                                                                                                      \\ \hline
        $BW$                                 & Bandwidth occupancy  
        
        \\ \hline
        $\alpha$                             & Transmitted symbol    
        
        \\ \hline
        $E_{\text{s}}$, $T_{\text{s}}$       &
        \begin{tabular}[c]{@{}c@{}}Energy per transmitted symbol,\\ Duration of the transmitted symbol\end{tabular}                                                                                                                                                                                                        \\ \hline
        $\mu$                                & ${2\pi}$ phase increment correcting factor                                                                                                                \\ \hline
        $d_{\text{min}}^2$, $d_{\text{B}}^2$ & \begin{tabular}[c]{@{}c@{}}Minimum squared Euclidean distance,\\  Minimum squared Euclidean distance upper\\  bound\end{tabular} \\ \hline
        $N$                                  & Number of observation symbols                                                                                                    \\ \hline
        $\gamma$                             & Difference data sequence                                                                                                         \\ \hline
        $P_{k}$                              & Priori probabilities of the data symbols                                                                                         \\ \hline
        $N_{\text{s}}$                       & \begin{tabular}[c]{@{}c@{}}Number of states needed to implement the\\  MLSD receiver\end{tabular}                                \\ \bottomrule
        \end{tabular}
    }
        \end{table}

The  SSB-FSK signal is defined as CPM \cite{anderson2013digital} with the complex representation 
\begin{align}
\label{eq:Modulation}
s(t,\alpha) &= \sqrt{\frac{E_{\text{s}}}{T_{\text{s}}}} \exp\Big\{j\phi(t;\alpha)\Big\},
\end{align}
where $E_{\text{s}}$ is the energy per transmitted symbol and $T_{\text{s}}$ is the duration of the symbol. The information-carrying phase is defined as
% \textcolor{red}{
% \begin{align*}
% \label{phi-alpha}
% \phi(t,\alpha) =  h \sum_{i=-\infty}^{+\infty} \alpha_i \phi_{0} (t-iT_{\text{s}}),
% \end{align*}}
\begin{equation}
    \phi(t,\alpha) = 2\pi \Tilde{h}\sum_{i=-\infty}^{+\infty} \alpha_i \phi_{0} (t-iT_{\text{s}})
\end{equation}
 where $\alpha_i$ is the transmitted symbol that takes values from the \textit{M-ary} level $0, 1,\ldots,(M-1)$. For instance, in order to preserve the SSB property, we can only use positive information symbols, responsible of the right side band spectrum. The negative values will introduce a lower side in the spectrum. Therefore, no antipodal coding is allowed~\cite{fares2018quantum,fares2017new}. Furthermore, $\tilde{h}=2h$, where $h$ is the modulation index used to ensure a $2\pi$ phase increment. The phase response $\phi_{0}(t)$ is represented as
% \textcolor{red}{\begin{align*}
% \phi_{0}(t)=\begin{cases}0 & t\leq{0}\\ \int_{-\infty}^{t} g(\tau)d\tau & {-LT_{\text{s}}/2}\leq t \leq{LT_{\text{s}}/2}, \end{cases}
% \end{align*}}

\begin{align}
    \begin{aligned}
    \phi_{0}(t) = \begin{cases}0 & t < 0\\{\frac{1}{4\pi}\int_{-\infty}^{t} }g(\tau)d{\tau} &0 \leq t < LT_{\text{s}} \\
    \frac{1}{2} & t\geq LT_{\text{s}}\end{cases}
    \end{aligned}
    \end{align}
where $g(t)$ is a Lorentzian frequency pulse truncated to a symbol duration  ${L>1}$ (\textit{partial-response}), defined as 

\begin{align} 
\begin{aligned}
\label{freqeuncy pulse}
g(t) = \frac{d\phi_{0}(t)}{dt} = \mu\frac{2w^{2}}{t^{2}+w^{2}},\\
t \in [-LT_{\text{s}}/2,LT_{\text{s}}/2].
\end{aligned}
\end{align}

The variable $w$ is the pulse width, a key tuning parameter that significantly impacts the performance, mainly the transmitted signal's spectral efficiency $s(t,\alpha)$~\cite{fares2018quantum,fares2017new}. 

\begin{figure}[t]
    \includegraphics[width=\columnwidth,center]{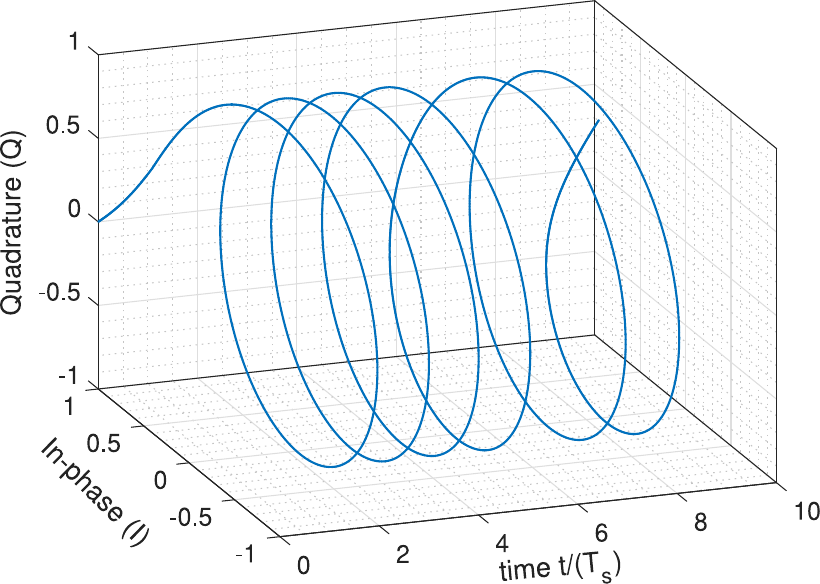}
    \caption{ Generated SSB-FSK signal with $L=4$, $w=0.3$ and $h=1$ of the bits sequence \{1, 1, 1, 1, 0, 1, 1, 0, 0, 1\}.}
    \label{fig:modualted_signal_complexe}
\end{figure}

\begin{figure}[t]
    \includegraphics[width=\columnwidth,center]{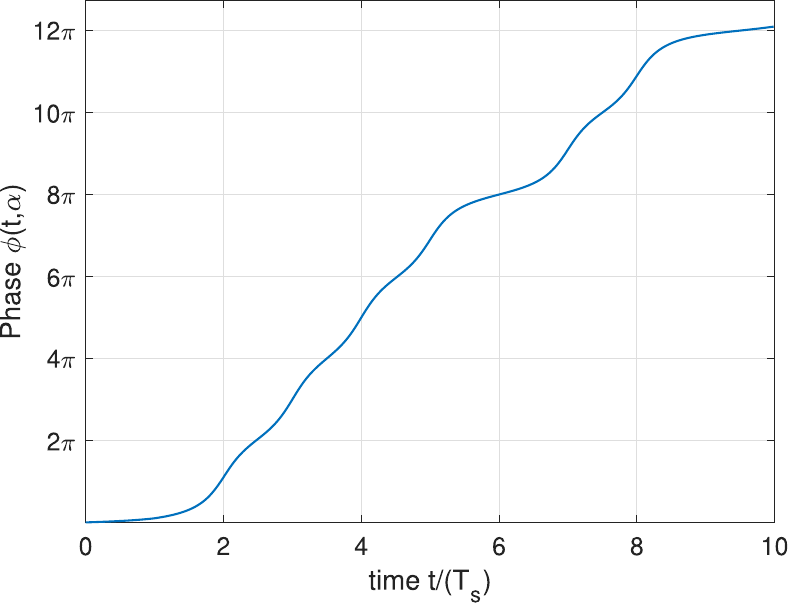}
    \caption{Evolution of the phase $\phi(t,\alpha)$ with $L=4$, $w=0.3$ and $h=1$ of the bits sequence \{1, 1, 1, 1, 0, 1, 1, 0, 0, 1\}.}
    \label{fig:modualted_signal_phase}
\end{figure}

Fig.~\ref{fig:modualted_signal_complexe} shows the constant envelope of the SSB-FSK modulated signal in the time domain; Fig.~\ref{fig:modualted_signal_phase} depicts the continuity of SSB-FSK phase.

As the Lorentzian pulse is characterized by a slow decrease, a pulse truncation is needed to get acceptable values of pulse lengths (regarding especially receiver complexity). As a consequence of this truncation, $\mu$ is introduced as a correcting factor to keep a ${2\pi}$ phase increment to sustain the SSB property. The correcting factor $\mu$ is defined as the ratio between the total phase increment without any truncation and the one obtained after Lorentzian truncation:
\begin{align}
\mu (L) = \frac{2\pi}{\int_{-LT_{\text{s}}/2}^{LT_{\text{s}}/2} \frac{2w^{2}}{t^{2}+w^{2}}dt} = \frac{\pi}{2\arctan(\frac{LT_{\text{s}}}{2w})}.
\end{align}

\begin{figure}[t]
    \includegraphics[width=\columnwidth,center]{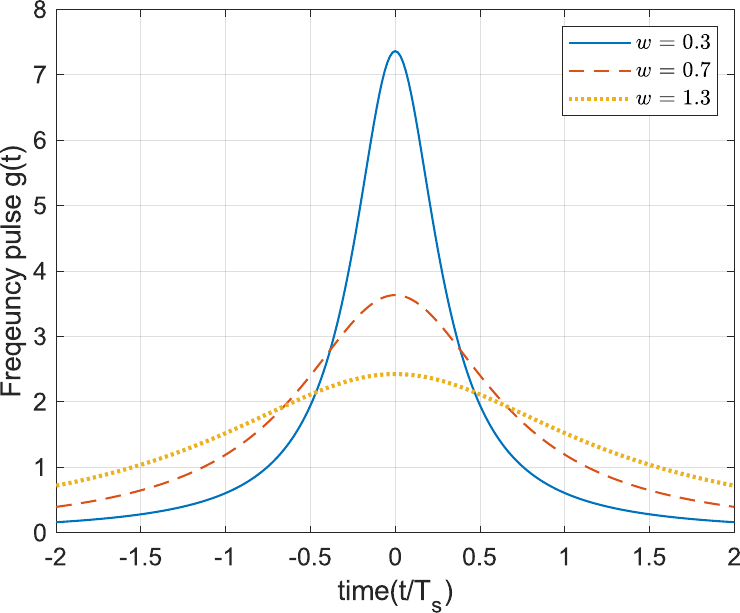}
    \caption{Lorentzian pulse (L=4) for different width values, $w=0.3,0.7,1.3$.}
    \label{fig:Lorent_w}
\end{figure}

The effect of $w$ on the frequency pulse $g(t)$ for $L=4$ is presented in Fig.\ref{fig:Lorent_w}, for different values of $w=0.3,0.7,1.3$. As shown in Fig.\ref{fig:Lorent_w}, it is obvious that increasing $w$ rises the tails of frequency pulse $g(t)$, which means an increase in inter-symbol interference (ISI).

\section{Single-Sideband CPM error probability}
\label{sec3}
In this section, we consider the performance of SSB-FSK when transmitted over an AWGN (additive white Gaussian noise) channel, assuming there is no channel coding. To evaluate the performance of the signal, we used the same method shown in~\cite[Ch.\ 3]{anderson2013digital}. We consider the union bound of the error probability, as the asymptotic performance for high values of $E_b/N_0$, where $E_b$ is the mean energy per information bit. The union bound is defined as in~\cite[Ch.\ 3, p.\ 55]{anderson2013digital}
\begin{align} 
\begin{aligned}
    \label{Prob_error}
P_e \sim Q\left(\sqrt{d_{\text{min}}^{2}\frac{E_b}{N_0}}\right),
\end{aligned}
\end{align}
where $d_{min}^2$ is the minimum squared Euclidean distance. The function $Q(.)$ is the error Gaussian function defined by
\begin{align}
\begin{aligned}
Q(x) = \int_{x}^{+\infty} \frac{1}{\sqrt{2\pi}}e^{-v^{2}/2}dv.
\end{aligned}
\end{align}

The normalized squared Euclidean distance ($d^{2}$) is the Euclidean difference between any two transmitted SSB-FSK data sequences ${\boldsymbol{\alpha}}$  and ${\boldsymbol{\alpha}'}$. The general form of the normalized squared Euclidean distance for any set of signals is given by
\begin{align}
\label{eq_dmin_1}
\begin{aligned}
d^{2}(\alpha,\alpha') = \frac{1}{2E_b} \int_{0}^{NT_{\text{s}}} |s(t,\alpha)-s(t,\alpha')|^2dt,
\end{aligned}
\end{align}
where $NT_{\text{s}}$ is the observation symbol intervals. The normalized squared Euclidean distance $d^{2}(\alpha,\alpha')$ can be represented as a function of a single difference data sequence $\boldsymbol{\gamma}$, where $\boldsymbol{\gamma} = \alpha - \alpha'$. Therefore, (\ref{eq_dmin_1}) can be rewritten as (see appendix A for the derivation details)\cite[ch.2]{anderson2013digital}
\begin{equation}
    \label{dmin}
d^{2}(\gamma) = \log_{2}M\Big\{N-\frac{1}{T_{\text{s}}} \int_{0}^{NT_{\text{s}}} \cos[\phi(t,\gamma_N)]dt\Big\},
\end{equation}
where $\phi(t,\gamma_N)$ is the phase difference trajectories. 

% \textcolor{red}{To retrieve the same initial assumption considered in the derivations presented in~\cite[Ch.\ 3]{anderson2013digital}, we rearrange (2) as
% \begin{equation*}
% \phi(t,\gamma) = 2\pi \Tilde{h}\sum_{i=-\infty}^{+\infty} \gamma_i \tilde{\phi_{0}} (t-iT_{\text{s}})
% \end{equation*}
% where ${\Tilde{h}=2h}$. The phase function $\Tilde{\phi_{0}}(t)$ is given by
% \begin{align*}
% \begin{aligned}
% \tilde{\phi_{0}}(t) = \begin{cases}0 & t < 0\\{\frac{1}{4\pi}\int_{-\infty}^{t} }g(\tau)d{\tau} &0 \leq t < LT_{\text{s}} \\
% \frac{1}{2} & t\geq LT_{\text{s}}\end{cases}
% \end{aligned}
% \end{align*}}

\begin{figure}[t]
\centering
    \includegraphics[width=\columnwidth]{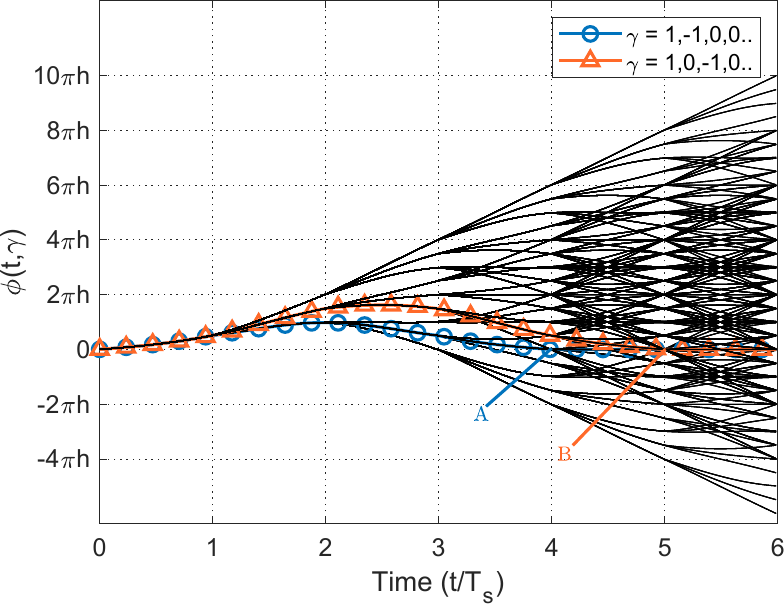}
    \caption{Phase difference tree for binary SSB-FSK with $L=3$ and $w=0.9$. A and B indicate merges.}
    \label{fig:phase_diff_tree}
\end{figure}

To calculate the minimum normalized squared Euclidean distance $(d_{\text{min}}^2)$ for $N$ observation symbols, we need to find pairs of $\gamma$ that minimize (\ref{dmin}). It is clear that (\ref{dmin}) has positive integral, which means that the higher the $N$ we observe, the higher the Euclidean distance we obtain. Consequently, if we obtain an upper bound for $N \rightarrow \infty$, we obtain an upper bound at any defined $N$. This suggests the search for the difference phase trajectories $\phi(t,\gamma)$ with the shortest duration time $\tau$, where $\phi(t,\gamma)$ is equal to zero at all times $t\ge\tau$. Hence, an upper bound $(d_{\text{B}}^2)$ on the Euclidean distance $d^{2}(\gamma)$ can be obtained using only $\phi(t,\gamma)$ with the shortest length $\tau$. To better clarify the idea, we use in Fig.~\ref{fig:phase_diff_tree} the difference phase tree for general modulation index $h$ , as an example to illustrate the method we used to obtain the boundary. Since we are presenting different phase trees, the first symbol $\gamma_{0}$ cannot be zero, since $\alpha_{0}$ and $\alpha_{0}'$ must not be the same.  Using the difference phase tree, we can evaluate $\gamma$ with the shortest length of $\tau$. We search for the merger points between two different paths using the different phase trajectories ($\phi(t,\gamma)$) that start at $t=0$, and coincide with the x-axis at a certain time $t =t_k$, and coincide all the time after that for all $t >t_k$. In Fig.~\ref{fig:phase_diff_tree}, we showed two of the merger points $A$ and $B$, and two of the best $\gamma$ combinations with the shortest $\tau$ (presented as marked lines). We can directly see that the difference phase tree with the best $\gamma$ combinations (presented as marked lines) coincide at $t$ equal to $4$ and $5$ with the x-axis respectively at the points $A$ and $B$, and continue to be aligned with the x-axis from this points to the end. Usually, the time instant of the first merger point is obtained at $t=(L+1)T_{\text{s}}$, for non-weak modulation schemes~\cite[Ch.\ 3, p.\ 73]{anderson2013digital}. The phase difference tree, with the best difference sequence $\gamma$ is defined by
\begin{align}
\begin{aligned}
    \label{eq:gamma}
\gamma_{i}= \begin{cases}0 & i < 0\\1,2,\ldots,(M-1),  &i=0\\
0,\pm1,\pm2,\ldots,\pm(M-1),& 0<i<m+1\\0 & i\ge m+1 \end{cases},
\end{aligned}
\end{align}
given that
\begin{align}
\begin{aligned}
\label{eq:sum0}
\sum_{i=0}^{m} \gamma_{i} = 0.
\end{aligned}
\end{align}

The variable $m$ is the index of the merger points; the higher the number of mergers {$m$} we select, the tighter upper bound $d_{\text{B}}^2$ we obtain (in Fig. \ref{fig:phase_diff_tree}, we showed only two mergers points $A$ and $B$ for $m=2$). 
% \textcolor{red}{However, it is not exactly known the numbers of mergers $m$ needed so we cannot tighten the upper bound $d_{\text{B}}^2$ anymore~\cite[Ch.\ 3, p.\ 74]{anderson2013digital}.}
 As the exact numbers of merger $m$ is not known \cite[Ch.\ 3, p.\ 74]{anderson2013digital}, $m$ is selected by trial and is increased until there is no change in the upper bound $d_{B}^2$, with the knowledge that, there is no case where mergers higher than $L$ is needed~\cite[Ch.\ 3, p.\ 74]{anderson2013digital}.

The calculation of $d_{\text{min}}^2$  is obtained using the sequential search algorithm proposed in~\cite{aulin1981continuous}. The search algorithm requires the upper bound $d_B^{2}$; the main objective of the upper bound $d_B^{2}$ is to accelerate the algorithm search. For instance, any pair of phase trajectories with the corresponding $\gamma_N$ for a specified $h$ and $w$, having a distance value defined in (\ref{dmin}) larger than $d_B^{2}$, will not be used again.
The steps to obtain the minimum normalized squared Euclidean distance $(d_{\text{min}}^2)$ are summarized below:
\begin{itemize}
\item \textbf{Step 1}: select the number of mergers $m$. 
% \textcolor{red}{(even if the number of mergers needed is not exactly known, there is no case where mergers higher than $L$ is needed~\cite[Ch.\ 3, p.\ 74]{anderson2013digital}).} 
\item \textbf{Step 2}: use exhaustive search to calculate the set $S_{\gamma}$ of difference sequences $\gamma$  according to (\ref{eq:gamma}) and (\ref{eq:sum0}).
\item \textbf{Step 3}: compute $d^2(\gamma)$ based on (\ref{dmin}) for each difference sequence obtained in the set $S_{\gamma}$.
\item \textbf{Step 4}: the upper bound of the minimum distance $d_{\text{B}}^2$ is equal to the minimum bound of all normalized squared Euclidean distance obtained from \textbf{step 3} ($d_{\text{B}}^2=\underset{\gamma \in S_{\gamma}}{min} \: d^{2}(\gamma)$).
\item \textbf{Step 5}: calculate $d_{\text{min}}^2$ from the algorithm presented in \cite{aulin1981continuous}, using the $d_{\text{B}}^2$ obtained from \textbf{step 4}.  
\end{itemize} 

\section{Single-Sideband CPM Power Spectrum}
\label{sec4}
The spectrum occupancy plays an important role in the transmitted signal's overall performance, as it is a limited resource. Therefore, it is important to minimize the usage of this resource as much as possible. In this section, we apply a numerical method to compute the power spectral density and, consequently, analyze the spectrum efficiency of the SSB-FSK signals for different parameters ($h, L,w, M$). The method is based on~\cite[Ch.\ 4]{anderson2013digital}~\cite{aulin1983easy} and~\cite[Ch.\ 3]{proakis2008digital}, which is a generalization of the solution developed by Greenstein for time-limited phase response CPM signals~\cite{greenstein1977spectra}. 

\subsection{Non-Integer Modulation Index}
The method used in~\cite[Ch.\ 4]{anderson2013digital}~\cite{aulin1983easy} and~\cite[Ch.\ 3]{proakis2008digital} for the power spectrum calculation is based on the evaluation of the autocorrelation of the Fourier transform function, given that $s(t,\alpha)$ is cyclostationary. Therefore, we first introduce the autocorrelation function, where we assume a baseband signal, given by

\begin{flalign} 
\begin{aligned}
    \label{autocorr}
R(\tau) &= R(\tau'+mT_{\text{s}}) \\ 
&=\frac{1}{T_{\text{s}}}\int_{0}^{T_{\text{s}}} \prod_{i=1-L}^{m+1}\Big\{\sum_{k=0}^{M-1} P_{k}\exp(j 2 \pi h k \\
&\quad[\phi_{0}(t+\tau'- (i-m)T_{\text{s}})-\phi_{0}(t-iT_{\text{s}})])\Big\}dt.
\end{aligned}
\end{flalign}
The variable $\tau = \tau' + mT_{\text{s}}$ is the time difference over the interval $[0,(L+1)T_{\text{s}}]$, with $0\leq\tau' <T_{\text{s}}$ and $m=0, 1, 2,\ldots$. The variable $P_{k}$ is the priori probability of the data symbol $k$.
The PSD is obtained by using the Fourier transform of (\ref{autocorr}), given by
\vspace{3ex}
\begin{equation} \label{PSD1}
\end{equation}
\vspace{-10ex}
\begin{flalign*} 
    \begin{aligned}
        S(f) &=2\:\Re \left\{\int_{0}^{LT_{\text{s}}} R(\tau)e^{-j2\pi f \tau}d\tau \parenthnewln
        + e^{-j2\pi f LT_{\text{s}}} \!\!\sum_{m=0}^{\infty}\!C_{\alpha}^{m}e^{-j2\pi f m T_{\text{s}}}\!\!\int_{0}^{T_{\text{s}}} \!\! R(\tau \! + \! LT_{\text{s}})e^{-j2\pi f \tau}d\tau \right \},
    \end{aligned}
\end{flalign*}
where 
\begin{equation}
    C_{\alpha}= \sum_{k=0}^{M-1} P_{k} e^{j h \pi k}
\end{equation}
for $h$ values for which $|C_{\alpha}|<1$. The summation in (\ref{PSD1}) converges, leading to 
\begin{equation}
    \label{Calpha1}
    \sum_{m=0}^{\infty}C_{\alpha}^{m}e^{-j2\pi f m T_{\text{s}}} = \frac{1}{1-C_{\alpha}e^{-j2\pi f T_{\text{s}}}}.
\end{equation}
Inserting (\ref{Calpha1}) into (\ref{PSD1}), we obtain the PSD expression given by
\vspace{3ex}
\begin{equation} \label{PSD2}
\end{equation}
\vspace{-10ex}
\begin{flalign*} 
    \begin{aligned}
        S(f) &= 2 \Re \left\{\int_{0}^{LT_{\text{s}}} R(\tau)e^{-j2\pi f \tau}d\tau \parenthnewln
        \quad + \frac{e^{-j2\pi f LT_{\text{s}}}}{1-C_{\alpha}e^{-j2\pi f T_{\text{s}}}} \int_{0}^{T_{\text{s}}} R(\tau + LT_{\text{s}})e^{-j2\pi f \tau}d\tau\right\}.
    \end{aligned}
\end{flalign*}
However, this PSD equation (\ref{PSD2}) is only valid for $|C_{\alpha}|<1$, i.e. for non-integer $h$. Therefore, in the Subsection~\ref{subB}, we apply certain modifications to obtain the PSD for integer $h$.
\subsection{Integer Modulation Index}
For $|C_{\alpha}|=1$ (integer $h$), (\ref{Calpha1}) is not valid anymore. However, with $|C_{\alpha}|=1$ we can set
\begin{equation}
    \label{Calpha0}
    C_{\alpha} = e^{j2\pi v}, \quad 0\leq v<1.
\end{equation} 
Based on (\ref{Calpha0}), the summation in (\ref{Calpha1}) can be rewritten as~\cite[Ch.\ 3, p.\ 141]{proakis2008digital}
\vspace{2.8ex}
\begin{equation} \label{Calpha2}
\end{equation}
\vspace{-10ex}
\begin{align*}
    \begin{aligned}
    &\sum_{m=0}^{\infty} e^{-j2\pi T_{\text{s}}(f-v/T_{\text{s}})m}= \frac{1}{2} \\
    & + \frac{1}{2T_{\text{s}}}\sum_{m=-\infty}^{\infty}\delta(f-\frac{v}{T_{\text{s}}}-\frac{m}{T_{\text{s}}})
    -j\frac{1}{2}\cot\pi T_{\text{s}}(f-\frac{v}{T_{\text{s}}}).
    \end{aligned}
\end{align*}
Inserting (\ref{Calpha2}) into (\ref{PSD1}) leads to the complete PSD expression for integer $h$, which covers discrete and continuous components of the spectrum. Moreover, based on (\ref{Calpha2}) the discrete spectrum spikes are located at frequencies 
\begin{align}
    f_{m} = \frac{m+v}{T_{\text{s}}}, \quad 0\leq v<1. \quad  m=0,1,2...
\end{align}
\label{subB}
\section{Multi-Objective Optimization}
\label{sec5}
SSB-FSK has several conflicting objectives that need to be optimized, where the major challenge is to find tradeoffs for designing a competitive SSB CPM system. We have thus divided this section into three parts:
\begin{itemize}
\item[--] Objective functions: in this part, we summarize the performance metrics considered in the optimization analysis.
\item[--] Design space: A massive number of configurations has to be tested due to the variety of the parameters. Therefore, it is interesting to reduce this number of combinations by excluding situations that cannot be envisaged, i.e., by defining a feasible region of the optimization problem variables.
\item[--] Optimization method: as we have several objectives to optimize, different methodologies are possible. One solution, among many, is to optimize one performance metric when considering others constrained to a certain threshold (drawn from well-known other CPM schemes). Using this methodology, we define plenty of optimization problems (as many as possible combinations between considered objective functions and fixed thresholds for constrained performance metrics). Therefore, the best course of action is to jointly optimize all performance metrics by defining several intermediate configurations leading to different tradeoffs depending on the order of priority given for these objective functions.
\end{itemize}
\subsection{Objectives Functions}  
\begin{itemize}
    \item \textit{Minimum Normalized Squared Euclidean Distance} ($d_{\text{min}}^{2}$): measures the energy consumption of the signals, based on the error probability bound given in (\ref{Prob_error}). It is clear that an increase of $d_{\text{min}}^{2}$ can be seen as a reduction in the mean energy per information bit $E_{b}$ for the same targeted probability error.
    \item \textit{Bandwidth Occupancy} ($BW$): estimates the $BW$ occupancy of the signal. We divided the $BW$ occupancy into two measurements; the first one computes $99\%$ of the signal power inside $BT_{\text{b}}$, where $T_{\text{b}}=T_{\text{s}}/\log_{2}M$ and $ B $ represents the total bandwidth occupancy and not half of it due to the unsymmetrical spectrum shape form of the SSB-FSK modulation. The second measurement computes $99.9\%$ of the power inside $BT_b$. With the second measurement, we obtain a better idea about the out-of-band leakage. We noted the two measurements $B_{99}$ and $B_{999}$, respectively. These measurements' objective is to estimate the $BW$ occupancy of $BT_{\text{b}}$. We note that the bandwidth $BT_{\text{b}}$ is normalized to the data rate in terms of bits carried per second so that schemes with different modulation levels $M$ can be compared.
    \item \textit{Complexity} ($N_{\text{s}}$): it is given by the number of states used to implement the MLSD receiver \cite[ch.3, p.111-p.113]{anderson2013digital}:
    \begin{align}
        \label{eq:Complexity}
        \begin{aligned}
            N_{\text{s}}=\begin{cases}  p & L=1 \\
                                        pM^{L-1} & L>1
            \end{cases}, 
        \end{aligned}
      \end{align}
    where $p$ is the number of phase states obtained from the modulation index $h = \frac{2m}{p}$ (for SSB-FSK, $\tilde{h}=\frac{2m}{p}$).
    \item \textit{Single-Sideband Loss} (SSB-LOSS): the SSB-LOSS is a specific objective function defined only for SSB-FSK signals (as they are the only CPM signals having this spectral feature). SSB-LOSS estimates the percentage of power loss in the lower band (the respectively upper band for negative modulation indices $h$) of the PSD.\\
\end{itemize}
Due to the enormous computational complexity of all of these metrics, we have to establish an order of priority order among them. Therefore, we provide the highest priority degree to the signal energy consumption, $BW$ occupancy and complexity, where we try to find certain tradeoffs between these three objective functions. Based on these tradeoffs, we subsequently try to reduce the SSB-LOSS. 

\subsection{Space Design}
The initial step for solving a multi-objective function is to define a feasible region for the problem. In this study, we are mainly interested in three objective functions: energy consumption, $BW$ occupancy, and complexity. Therefore, the feasible region for the constraints is discussed hereafter :
\begin{enumerate}
    \item \textit{Modulation index} $h$: it is important to keep a small $h$, where large $h$ has huge impact on the occupied $BW$~\cite[Ch.~4]{anderson2013digital} \cite{Yang2013,Messai2016} (more information about the effect of $h$ on the spectrum of SSB-FSK signals, are presented in~\ref{subsec2}). Therefore, we define the modulation index range $0.01 \leq h \leq 2$, with an increment step size $h_{\mathrm{s}}=0.01$. 
    \item \textit{Pulse Length} $L$: an increase in $L$ will produce an increase in complexity (\ref{eq:Complexity})~\cite[ch.~4, p.~248]{proakis2008digital}. Therefore, in the literature, for all CPM modulation schemes, it is difficult to find a CPM design with $L>8$~\cite{anderson2013digital,syed2007comparison}\cite{Perrins2005}, where $L=8$ is already a complex system. However, in~\cite{fares2018quantum}, we used a pulse length $L=12$. Therefore, we define the pulse length range as $1 \leq L \leq 12$ with increment step size $L_{\mathrm{s}}=1$. We will not exceed this maximum value of $L=12$ to not explode the receiver complexity gaining very little on the SSB property.
    \item  \textit{Modulation level} $M$: an increase in $M$ will create an increase in complexity (\ref{eq:Complexity})~\cite[ch.~4, p.~248]{proakis2008digital}, and degradation in $BW$ occupancy $BT_{\text{b}}$ for the same modulation index $(h)$~\cite{anderson2013digital}. Therefore we selected $M$ to take only discrete values $\{2,4,8\}$.
    \item \textit{Pulse Width} $w$: $w$ is a new parameter introduced by the SSB-FSK. Based on~\cite{fares2018quantum}, the only idea we have about $w$ is that increasing $w$ will impact the SSB-FSK PSD exponential decrease, which can be quantified to $\exp(−4 \pi w/T_b)$ for the particular case $h=1$. Therefore, it is interesting to study the effect of the pulse width $w$ more deeply on both energy computation and $BW$ occupancy $BT_{\text{b}}$. To fully cover the effect of the pulse width $w$, we need to define a numerical range that covers up the \textit{useful} region (a reduced set into the \textit{feasible} region). The pulse width $w$ can only be found in the frequency pulse given by (\ref{freqeuncy pulse}). It is clear from this expression that no restrictions are found to delimit possible values of $w$ (expect that it has to be a non-null positive real). Therefore, to cover the full \textit{feasible} region, we should take into consideration all values of pulse width $w$, i.e., $w  \in \: ]0,\infty[$. Consequently, we define $w_{\text{lim}}$ a new limit on the pulse width $w$, where for $w_{\text{lim}}\leq w \leq \infty$ we obtain approximately the same frequency pulse $g(t)$, hence, it will be possible to reduce the study to $w\in \: ]0,w_{\text{lim}}]$, which is for us now the \textit{useful} region. The calculation of $w_{\text{lim}}$ is straightforward using the \textit{2-norm} defined as
    \begin{align}
        \label{NORM2}
        ||x||_{2} = \sqrt{\sum_{i}^{} |x_{i}|^2}.
    \end{align}
    The error is then calculated using
    \begin{align}
        \label{error}
        \epsilon=\frac{||\hat{x}-x||_{2}}{||x||_{2}},
    \end{align}
    where $x$ is the frequency pulse $g_{\infty}(t)$, for $w=\infty$ (in simulation we take $w=1000$). Furthermore, $\hat{x}$ is the estimated frequency pulse $g_{\text{lim}}(t)$, for $w=w_{\text{lim}}$. We calculate $w_{\text{lim}}$ for each $L$ in the defined interval. The criteria to obtain $w_{\text{lim}}$ from (\ref{error}) is to maintain an error $\epsilon \leq 10^{-1}$. In Table~\ref{Tab:WOPT}, we present $w_{\text{lim}}$ with respect to the pulse length $L$, for binary SSB-FSK. Finally, we define the pulse width range $0.1 \leq w \leq w_{\text{lim}}$. It is worth noting that $w_{\text{lim}}$ is increasing with respect to the pulse length $L$. Therefore, more configurations have to be considered when increasing $L$ as the defined range of $w$ is getting wider. Moreover, the increment step size is $w_{\text{s}}=0.1$.
\end{enumerate} 

\begin{table}[t]
    \centering
    \captionsetup{justification=centering}
    \caption{\\Pulse width limit $w_{\text{lim}}$ with respect to pulse length $L$ for binary SSB-FSK.}
    \label{Tab:WOPT}
    \begin{tabular}{@{}ccccccc@{}}
    \toprule
    $L$ & 2 & 4 & 6 & 8 & 10 & 12 \\ \midrule
    $w_{\text{lim}}$ & 1.6 & 3.2 & 4.8 & 6.4 & 7.9 & 9.5 \\ 
    \bottomrule
    \end{tabular}
\end{table}

\begin{figure}[t]
        \centering
            \includegraphics[width=\columnwidth]{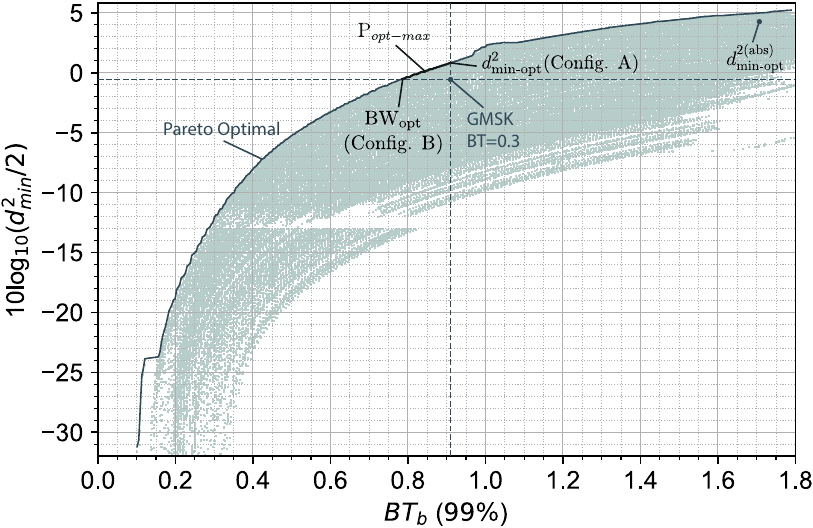}
            \caption{Two objective functions Pareto optimum plot for binary $L$SSB-FSK using $B_{99}$. GMSK for $BT=0.3$ is presented as a reference point.}
            \label{fig:Pareto optimum M2_99}
        \end{figure}
\begin{table}[]
    \centering
    \captionsetup{justification=centering}
    \caption{\\SSB-FSK optimum minimum normalized squared Euclidean distance $d_{\text{min}-\text{opt}}^2$ and optimum bandwidth occupancy $BW_{\text{opt}}$ for $99\%$ bandwidth BW occupancy, for different modulation levels $M$.}
    \resizebox{0.9\columnwidth}{!}{%
    \begin{tabular}{@{}cccccccc@{}}
    \toprule
    \multicolumn{8}{c}{\boldmath$d_{\text{min}-\text{opt}}^2$ - Config.~A}                   \\ \midrule \midrule
    $M$ & $L$ & $w$ & $h$  & $d_{\text{min}}^2$ & $BW$  & SSB-LOSS (\%) & $N$ \\ \midrule
    2   & 5   & 1.3 & 0.78 & 2.4         & 0.906 & 1.764     & 15  \\ \midrule
    4   & 2   & 0.7 & 0.49 & 3.53        & 0.906 & 2.561     & 25   \\ \midrule
    8   & 2   & 0.6 & 0.36 & 3.25        & 0.904 & 1         & 8     \\ \bottomrule
    \multicolumn{8}{c}{\boldmath$BW_{\text{opt}}$ - Config.~B}                           \\ \midrule \midrule
    $M$ & $L$ & $w$ & $h$  & $d_{\text{min}}^2$ & $BW$  & SSB-LOSS (\%) & $N$ \\ \midrule
    2   & 5   & 1.2 & 0.65 & 1.774       & 0.785 & 2         & 14  \\ \midrule
    4   & 2   & 0.8 & 0.33 & 1.773       & 0.65  & 2.732     & 10   \\ \midrule
    8   & 2   & 0.6 & 0.26 & 1.8         & 0.677 & 1.21      & 8     \\ \bottomrule
    \end{tabular}
    }
    \label{Tab:opt_99}
    \end{table}
    \subsection{Optimization Method (Without Complexity)}
    \label{sec:opt_1}
    In multi-objective problems, there is no single global optimal solution. Therefore, \textit{Pareto optimum} is used to find all optimal points. A solution is said to be \textit{Pareto optimum} ($P_{\text{opt}}$), if there is no other solution that can improve at least one of the objective functions without reducing the other objective functions. The objective functions in our case are
    \begin{align}
    F(x) = [F_{1}(x), F_{2}(x)],
    \end{align}
    where
    \begin{align}
    F(x_{1})>F(x_{2}) \iff \forall_{i} \: F_{i}(x_{1})>F_{i}(x_{2}). 
    \end{align}
    $F_{1}(x)$ and $F_{2}(x)$ are the $10\text{log}_{10}(d_{\text{min}}^2/2)$ and the inverse of the bandwidth occupancy ($1/BT_{\text{b}}$) respectively. The constraints variables ($h,L,M,w$) are denoted here by $x \in \mathbb{R}^{4}$.
    The definition of \textit{Pareto optimum} is given by (we present the weak \textit{Pareto optimum}) :
     \begin{flalign}
         \begin{aligned}
             x^{*} = P_{\text{opt}} \iff \nexists \: y \: \text{such that} \: F(y)>F(x^{*}).
         \end{aligned}
     \end{flalign}
 For more information about the \textit{Pareto optimum}, please refer to~\cite{marler2004survey,ngatchou2005pareto}.
 
 We used a \textit{Brute Force} method to obtain all the values of $F_{1}(x)$ and $F_{2}(x)$ for the space $Z$, where $Z$ is constrained by the parameters $h,L,M,w$ and their defined ranges. Based on this \textit{Brute Force}, we apply the \textit{Pareto optimum} optimization to obtain all the optimum solutions for $Z$. 

 \textit{Remarks}:
 \begin{itemize}
     \item We calculate the minimum normalized squared Euclidean distance $(d_{\text{min}}^2)$ using the algorithm introduced in Section~\ref{sec3}, where we consider all the mergers for merger index $m=3$ (the number of mergers $m$ found by trials). 
    %  \textcolor{red}{We consider only the first $4$ mergers ($m=3$) at $t=LT_{\text{s}}$, $t=(L+1)T_{\text{s}}$,\ldots,$t=(L+3)T_{\text{s}}$, to make sure we obtain a tightened upper bound $d_{\text{B}}^2$}.
      Moreover, we also consider $N_{\text{max}}=30$, where $N$ is the number of observation symbols.
     \item We calculate $B_{99}$ and $B_{999}$ $BW$ occupancy ($BT_{\text{b}}$) using the method presented in Section~\ref{sec4}. Hence, we obtain different $P_{\text{opt}}$ for each bandwidth $BW$ occupancy $BT_{\text{b}}$ measures.
 \end{itemize}

\begin{figure}[t]
        \centering
            \includegraphics[width=\columnwidth]{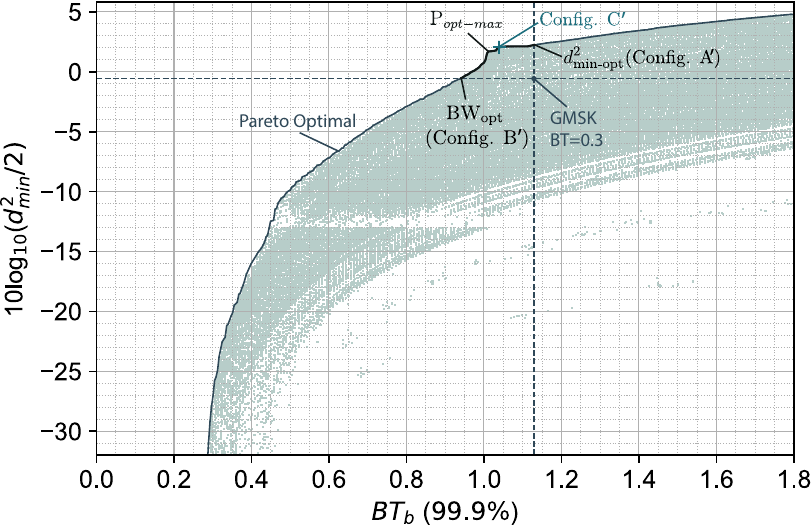}
            \caption{Two objective functions Pareto optimum plot for binary $L$SSB-FSK using $B_{999}$. GMSK for $BT=0.3$ is presented as a reference point.}
            \label{fig:Pareto optimum M2_999}
        \end{figure}
\begin{table}[]
    \centering
    \captionsetup{justification=centering}
    \caption{\\SSB-FSK optimum minimum normalized squared Euclidean distance $d_{\text{min}-\text{opt}}^2$ and optimum bandwidth occupancy $BW_{\text{opt}}$ for $99.9\%$ bandwidth $BW$ occupancy, for different modulation levels $M$.}
    \resizebox{0.9\columnwidth}{!}{%
    \begin{tabular}{@{}cccccccc@{}}
    \toprule
    \multicolumn{8}{c}{\boldmath$d_{\text{min}-\text{opt}}^2$ - Config.~A$'$}                   \\ \midrule \midrule
    $M$ & $L$ & $w$ & $h$  & $d_{\text{min}}^2$ & $BW$  & SSB-LOSS (\%) & $N$ \\ \midrule
    2   & 12  & 0.8 & 1.04 & 3.346       & 1.129 & 0.366         & 18  \\ \midrule
    4   & 2   & 0.7 & 0.44 & 2.98        & 1.25  & 2.611         & 7   \\ \midrule
    8   & 2   & 0.7 & 0.35 & 3.025       & 1.114 & 0.955         & 8    \\ \bottomrule
    \multicolumn{8}{c}{\boldmath$BW_{\text{opt}}$ - Config.~B$'$}                           \\ \midrule \midrule
    $M$ & $L$ & $w$ & $h$  & $d_{\text{min}}^2$ & $BW$  & SSB-LOSS (\%) & $N$ \\ \midrule
    2   & 6   & 1.1 & 0.67 & 1.773       & 0.941 & 1.683         & 14   \\ \midrule
    4   & 2   & 0.7 & 0.33 & 1.814       & 0.902 & 2.930         & 8     \\ \midrule
    8   & 2   & 0.6 & 0.26 & 1.8         & 0.902 & 1.214         & 8      \\ \bottomrule
    \end{tabular}
    }
    \label{Tab:opt_999}
    \end{table}

Figs~\ref{fig:Pareto optimum M2_99} and \ref{fig:Pareto optimum M2_999} present the \textit{Pareto optimum} for binary $L$SSB-FSK \footnote{$L$SSB-FSK: $L$SSB-FSK presents the pulse length $L$ of SSB-FSK signal pulse, e.g., 3SSB-FSK is the SSB-FSK modulation for pulse length $L=3$.} using $B_{99}$ and $B_{999}$, respectively. Moreover, the GMSK performance for $BT=0.3$ is considered here as a reference point. Hence, for this part, we consider only the Pareto front values which have a bandwidth $BW$ occupancy that is less than or equal to the GMSK bandwidth $BW$ occupancy, and a minimum normalized squared Euclidean distance $d_{\text{min}}^2$ greater than or equal to the GMSK distance $d_{\text{min}}^2$. These values are illustrated in the upper left quarter of the plane delimited by the axis passing by the GMSK reference point in Figs \ref{fig:Pareto optimum M2_99} and \ref{fig:Pareto optimum M2_999}. We will refer to these values as $P_{\text{opt}-\text{max}}$. In Fig.~\ref{fig:Pareto optimum M2_99}, $P_{\text{opt}-\text{max}}$ is achieved with $26$ different configurations (parameters combinations). In Fig.~\ref{fig:Pareto optimum M2_999}, $P_{\text{opt}-\text{max}}$ is achieved with $29$ different configurations. Note that each point of $P_{\text{opt}-\text{max}}$ corresponds to a particular configuration, and the number of configurations depends on the parameters step used in the search ($L_{\text{s},W_{\text{s}}},h_{\text{s}}$).

\begin{figure}[t]
    \centering
        \includegraphics[width=\columnwidth]{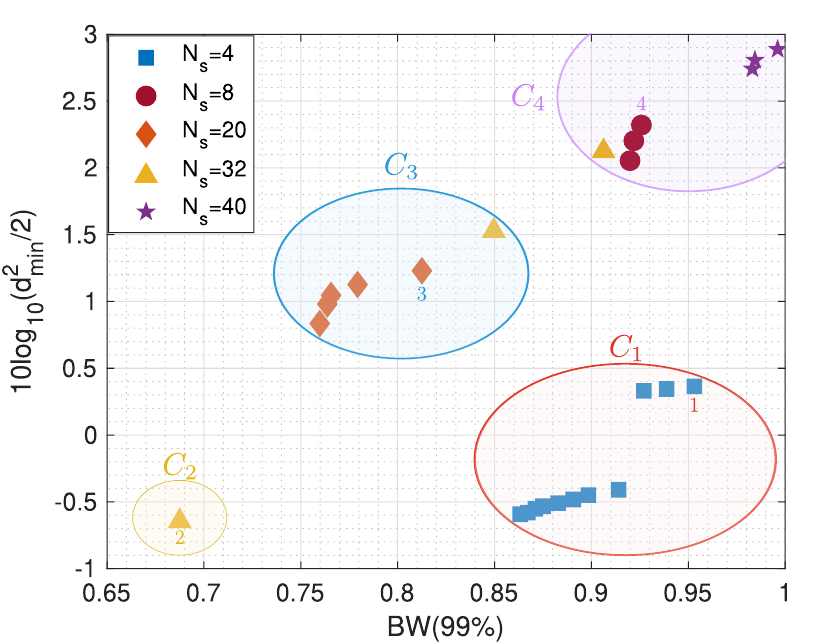}
        \caption{Three Pareto optimum objective functions configurations for binary $L$SSB-FSK using $B_{99}$.}
        \label{fig:Popt_3_99}
    \end{figure}

Tables \ref{Tab:opt_99} and~\ref{Tab:opt_999} present the optimum minimum normalized squared Euclidean distance $d_{\text{min}-\text{opt}}^2$ and optimum bandwidth occupancy $BW_{\text{opt}}$ for $99\%$ and $99.9\%$ occupancies, respectively, as function of the SSB-FSK parameters for different modulation levels $M$. Considering only the $P_{\text{opt}-\text{max}}$ configurations; $d_{\text{min}-\text{opt}}^2$ and $BW_{\text{opt}}$ are  the highest minimum normalized squared Euclidean distance $d_{\text{min}}^2$ (higher $d_{\text{min}}^2$ mean lower energy consumption) and the lowest bandwidth ($BW$) occupancy values obtained respectively. These points are denoted by Config.~A (for $d_{\text{min}-\text{opt}}^2$) and Config.~B (for $BW_{\text{opt}}$) for $B_{99}$ (Config.~A' and Config.~B' according to $B_{999}$). 
Besides, we also show the number of observation symbol intervals $N$ needed (the Viterbi algorithm memory highly depends on $N$), and the SSB-LOSS generated. The bandwidth occupancy $BW$ presented in the tables are normalized to the modulation level $M$. The \emph{Pareto optimum} analysis (Fig.~\ref{fig:Pareto optimum M2_99} and~\ref{fig:Pareto optimum M2_999}) gives all possible tradeoffs combining both normalized minimum Euclidean distance $d_{\text{min}}^2$ (y-axis, $10\text{log}_{10}(d_{\text{min}}^2/2)$) and bandwidth occupancy metrics (x-axis); tables \ref{Tab:opt_99} and~\ref{Tab:opt_999} give absolute optimum values, performing better than the GMSK, for each performance metric \emph{alone} and the resulting effect on the other one. 
Furthermore, in Fig.~\ref{fig:Pareto optimum M2_999}, we can see that the normalized minimum Euclidean distance $d_{\text{min}}^2$ performance has a small variation between Config.~A$'$ and Config.~C$'$ (the line is quite horizontal). Therefore, Config.~C$'$ seems an interesting point (another good tradeoff), since it offers almost the same normalized minimum Euclidean distance $d_{\text{min}}^2$ while decreasing significantly the bandwidth $BW$ occupancy. Config.~C$'$ corresponds to the following parameters combination ($L=12$, $w=0.7$, $h=0.99$); leading to $d_{\text{min-C}}^2=3.216$, $BW_{\text{\tiny{C}}}=1.043$.

    \begin{figure}[t]
        \centering
            \includegraphics[width=\columnwidth]{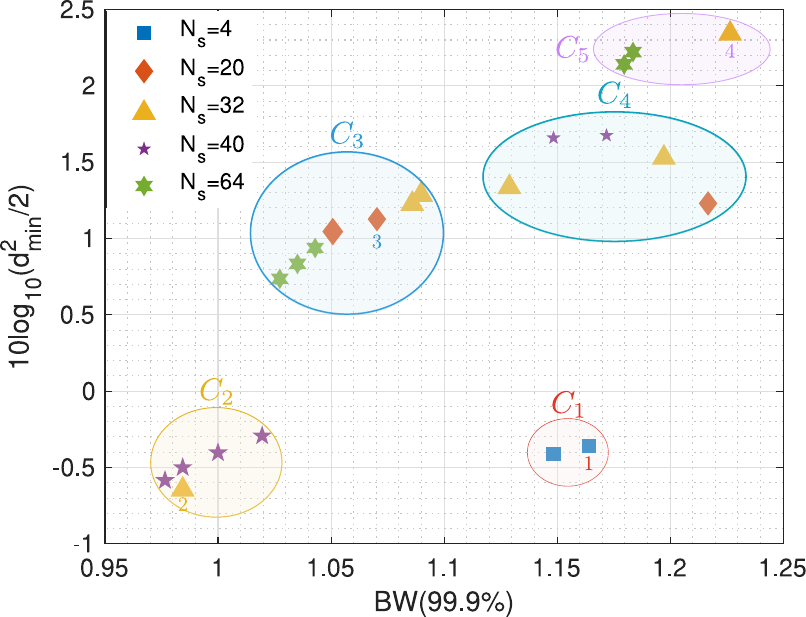}
            \caption{Three Pareto optimum objective functions configurations for binary $L$SSB-FSK using $B_{999}$.}
            \label{fig:Popt_3_999}
        \end{figure}

\subsection{Optimization Method (With Complexity)}
\label{sec:opt_2}
In this part, we consider three objective functions for optimization: $10\text{log}_{10}(d_{\text{min}}^2/2)$, the inverse of the bandwidth occupancy ($1/BT_{\text{b}}$) and the number of states ($N_{\text{s}}$). Similar to Section \ref{sec:opt_1}, we apply the $Pareto$ $optimum$ optimization with the three selected objective functions to obtain all the optimum solutions for the space $S$. Unlike $Z$, the space $S$ presents only the configurations with number of states $N_{\text{s}}\le64$ ($N_{\text{s}}>64$ high complexity). Moreover, the configurations in space $S$ can have an energy ($10\text{log}_{10}(d_{\text{min}}^2/2)$) values higher or equal to the GMSK energy minus ``$0.1$", and a bandwidth occupancy $BW$ lower or equal to GMSK $BW$ plus ``$0.1$".\\
In Fig.\ref{fig:Popt_3_99} and Fig.\ref{fig:Popt_3_999}, we illustrate the configurations obtained from the three objectives \textit{Pareto Optimum} optimization respectively for $B_{99}$ and $B_{999}$. Each configuration is presented by the $x$-axis (the $BW$ occupancy), the $y$-axis (the energy ($10\text{log}_{10}(d_{\text{min}}^2/2)$)), and the $z$-axis (the complexity ($N_{s}$)). The $z$-axis is illustrated using markers, where each different marker present a $N_{\text{s}}$ value. Since we have a large number of configurations ($25$ for $B_{99}$ and $22$ for $B_{999}$) and some of the configurations have similar performance (energy and $BW$). We decide to divide the configurations into clusters based on their performance similarity, and then from each cluster, we selected only one configuration. To obtain the clusters, we applied the \textit{k-means} algorithm [ch.14, p.509]~\cite{hastie2009elements} with Euclidean distance as cost function on the configurations obtained, taking into consideration only the energy and $BW$ occupancy of each configuration. Using the \textit{k-means} we were able to obtain $4$ and $5$ different cluster respectively for $B_{99}$ and $B_{999}$. Based on the clusters obtained, we select one configuration that represents the cluster based on two conditions:
\begin{enumerate}
    \item Since all configurations in the same cluster have approximately similar performance; we prioritized the complexity ($N_{s}$). Therefore, we select the configuration with the lowest complexity as the cluster representative.
    \item If two or more configurations in the same cluster have the same minimum complexity ($N_{s}$), we select the configuration with the maximum performance, measured as $10\text{log}_{10}(d_{\text{min}}^2/2)-$BW.
\end{enumerate}
In Fig.\ref{fig:Popt_3_99} and Fig.\ref{fig:Popt_3_999}, the selected configurations are presented with an index number. The index number presents the number of the row in Table.\ref{Tab:New_conf_99} and Table.\ref{Tab:New_conf_999} respectively for $99\%$ and $99.9\%$ $BW$ occupancy. These tables illustrate the parameters and performance of the selected configurations. The results from these tables will be discussed in Section \ref{sec6}, for the sake of comparison as they are presented alongside other relevant results from other known CPM schemes.\\
In Fig.\ref{fig:Popt_3_999}, we can see that we select only $4$ configurations out of $5$ because, compared to the configuration to be selected form the fourth cluster, the configuration number 3, while offering almost the same performance in terms of energy (a small difference of 0.1 dB), it is remarkably outperforming in terms of BW occupancy. Therefore, the configuration to be selected for cluster 4 is then ignored.  

\begin{figure}[t]
\centering
    \includegraphics[width=\columnwidth]{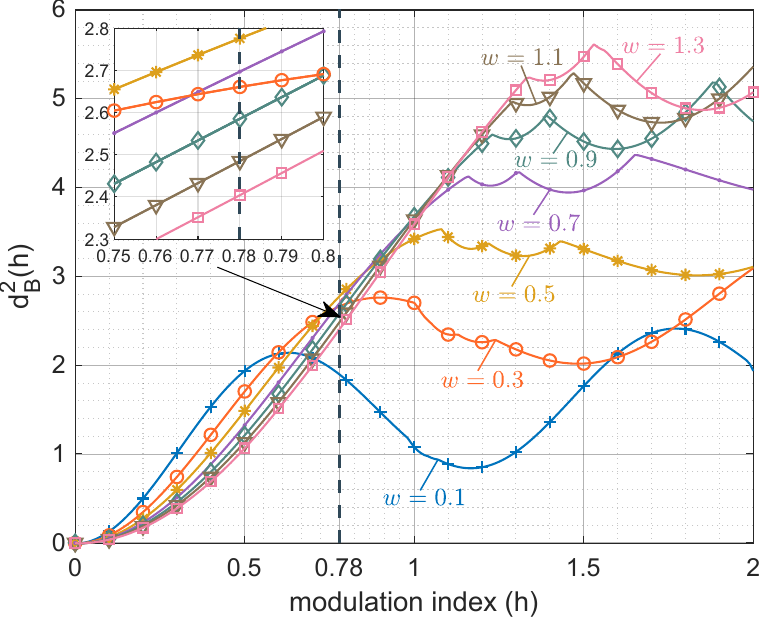}
    \caption{Effect of pulse width $w$ on the upper bound ($d_{\text{B}}^2$) as function of modulation index $h$ for binary 5SSB-FSK.}
    \label{fig:w_db}
\end{figure}

\section{Simulation Results}
\label{sec6}
In this section, we present the results obtained based on Section \ref{sec3} and \ref{sec4}. As this section is dedicated to illustrating SSB-FSK performance considering many metrics varying many parameters, we have chosen to do it subsequently for the sake of simplicity. Indeed, we first evaluate each metric apart and quantify the impact of all parameters on it: error probability performance in terms of $d_{\text{min}}$ are reported in sub-section \ref{subsec1}, BER simulation is illustrated in sub-section \ref{subsecBER} to verify the effectiveness of $d_{min}^2$ by illustrating the bit error performance of SSB-FSK signals, and power spectrum performance is depicted in subsection \ref{subsec2}. Second, we start to stack metrics together and evaluate performance when considered jointly: The interplay between $d_{\text{min}}^{2}$ and $BW$ is presented in sub-section \ref{subsec3}. By taking further into account the complexity, new tradeoffs are revealed in sub-section \ref{subsec3}. Finally, when synchronization advantage is suggested through the use of exclusively integer modulation indices, they are given in sub-section \ref{sec:advantage_synchro}, demonstrating the superiority of the SSB-FSK scheme among well-known CPM schemes.

\begin{figure}[t]
\centering
    \includegraphics[width=\columnwidth]{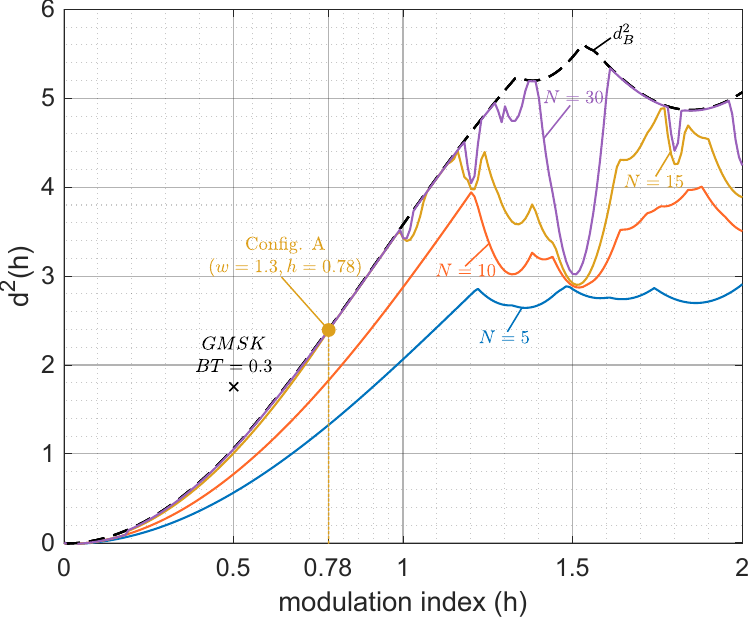}
    \caption{Minimum normalized squared Euclidean $d_{\text{min}}^2$ distance as function of modulation index $h$ for binary 5SSB-FSK with $w=1.3$.}
    \label{fig:h_dmin_w103}
\end{figure}   
    
\subsection{SSB-FSK CPM error probability Performance}
\label{subsec1}
The author in~\cite[Ch.\ 3]{anderson2013digital} describes, the effect of the parameters $h, M, L$ on $d_{\text{min}}^2$ for a variety of CPM schemes; the SSB-FSK has a quite similar scheme compared to the ones presented in~\cite[Ch.\ 3]{anderson2013digital} (\emph{unlimited time} CPM schemes with a symmetrical form with respect to zero). However, the pulse width $w$ is a specific parameter introduced particularly for SSB-FSK signals. Therefore, it is interesting to study its effects on the $d_{\text{min}}^2$.\\

Fig.~\ref{fig:w_db} depicts the variations of the upper bound $d_{\text{B}}^2$ as a function of modulation index $h$ for binary 5SSB-FSK, for pulse width $w=0.1,0.3,0.5,0.7,0.9,1.1,1.3$. Consequently, we are able to quantify the double effect of both parameters $h$ and $w$ on the error probability performance of SSB-FSK, which are presented in three points:
\begin{itemize}
    \item We obtain a better upper bound $d_{\text{B}}^2$ using a low pulse width $w$, e.g., $w=0.1$ for low modulation index region $h$, e.g., for $h\approx0.5$. Therefore, for small modulation index $h$, it is always better to use small pulse width $w$.
    \item We show that for higher modulation index $h$, a higher pulse width $w$ is preferred, e.g., for $h\approx1.5$ we obtain the best $d_{\text{B}}^2$ for $w=1.3$. This behavior is very similar to the effect of increasing the pulse length $L$, which is explained in details in~\cite[ch.3, p.75-76]{anderson2013digital}.
    \item In contrast to the second point, it is not always better to use higher pulse width $w$ for higher modulation index $h$. Since we notice that $d_{\text{B}}^2$ with $w=0.9$ is better than $w=1.3$, for the same $h\approx1.9$.
\end{itemize}

Based on the previous points, we cannot draw a general trend for the effect of the pulse width $w$ on the upper bound $d_{\text{B}}^2$ performance.\\
Besides, the impact of pulse width $w$ on the upper bound $d_{B}^{2}$ can be seen directly from Fig.~\ref{fig:Lorent_w}, where it is evident that the tails of the frequency pulse $g(t)$ goes up with the increase of pulse width $w$, which means an increase in inter-symbol interference (ISI). Fortunately, when the ISI is appropriately compensated at the receiver side, this can increase the redundant information (increase in memory effect). Hence, this introduces sizable gains in the upper bound $d_{B}^{2}$ values.
% Based on Fig.~\ref{fig:w_db}, we obtain a better upper bound $d_{\text{B}}^2$ using a low pulse width $w$, e.g., $w=0.1$ for low modulation index region $h$, e.g., for $h\approx0.5$. Therefore, for small modulation index $h$, it is always better to use small pulse width $w$. 
% In addition, we also show that for higher modulation index $h$, a higher pulse width $w$ is preferred, e.g., for $h\approx1.5$ we obtain the best $d_{\text{B}}^2$ for $w=1.3$. This behavior is very similar to the effect of increasing the pulse length $L$, which is explained in details in~\cite[ch.3, p.75-76]{anderson2013digital}. 
% However, it is not always better to use higher pulse width $w$ for higher modulation index $h$. Since we notice that the $d_{\text{B}}^2$ with $w=0.9$ is better than $w=1.3$, for the same $h\approx1.9$. 
% Therefore, we cannot conclude a general trend for the effect of the pulse width $w$ on the upper bound $d_{\text{B}}^2$ performance. 
% Moreover, the effect of pulse width $w$ on the upper bound $d_{B}^{2}$ can be seen directly from Fig.~\ref{fig:Lorent_w}, where it is obvious that the tails of the frequency pulse $g(t)$ goes up with the increase of pulse width $w$, which means an increase in inter-symbol interference (ISI). Fortunately, when the ISI is properly compensated at the receiver side, this can be seen as an increase on the redundant information (increase in memory effect), and hence this introduces sizable gains in the upper bound $d_{B}^{2}$ values. 

Fig.~\ref{fig:h_dmin_w103} illustrates the performance of minimum normalized squared Euclidean distance $d_{\text{min}}^2$ as function of $h$ for binary 5SSB-FSK with pulse width $w=1.3$. This configuration ($L=5$ and $w=1.3$) is chosen related to the results of the \emph{Pareto optimum} analysis given in the previous Section (it is Config.~A given also in Table~\ref{Tab:opt_99} row 1 relative to the binary scheme). Moreover, the results are illustrated for different observation symbol intervals $N=5,10,15,30$. In addition, the GMSK is also shown as x mark, and the $d_{\text{min}-\text{opt}}^2$ for binary $B_{99}$ SSB-FSK (Table.~\ref{Tab:opt_99} row $1$ or Config.~A in Fig.~\ref{fig:Pareto optimum M2_99}) is also shown as reference point. Fig.~\ref{fig:h_dmin_w103} shows that the most significant $d_{\text{min}}^2$ value slightly higher than $5$ can be reached, for $h=1.61$ and $N=30$: this is almost the absolute optimum without considering any other metric, and it is denoted by $d_{\text{min-opt}}^{2\:\text{(abs)}}$ (this is visible in Fig.~\ref{fig:Pareto optimum M2_99}, it is one of the extreme points at the upper right quarter of the plane). Compared to GMSK, we obtained a gain of $\approx 4.83$ dB. Note that this configuration, leading to $d_{\text{min-opt}}^{2\:\text{(abs)}}$, offers very poor $BW$ performance (this is also visible in Fig.~\ref{fig:Pareto optimum M2_99}). That is why in Table.~\ref{Tab:opt_99}, we have chosen the modulation index $h=0.78$ (Config.~A), which is shown as a reference point in the figure. This result, which is different from $d_{\text{min-opt}}^{2\:\text{(abs)}}$, is obtained because we are considering the bandwidth $BW$ occupancy alongside the minimum normalized squared Euclidean distance $d_{\text{min}}^2$. We can notice the decrease of modulation index $h$ (from $h=1.61$ to $h=0.78$), resulting in smaller bandwidth $BW$ occupancy. As a consequence, the energy consumption gain compared to GMSK decreases to $1.36$ dB. Besides, we can observe that, for certain specific modulation index $h$, the $d_{\text{min}}^2$ cannot reach the upper bound $d_{\text{B}}^2$, specially for modulation index $h=3/2$. This effect appears for all CPM schemes, and it is called weak modulation indices~\cite{anderson2013digital,aulin1981continuous}.  

\begin{figure}[t]
    \includegraphics[width=\columnwidth,center]{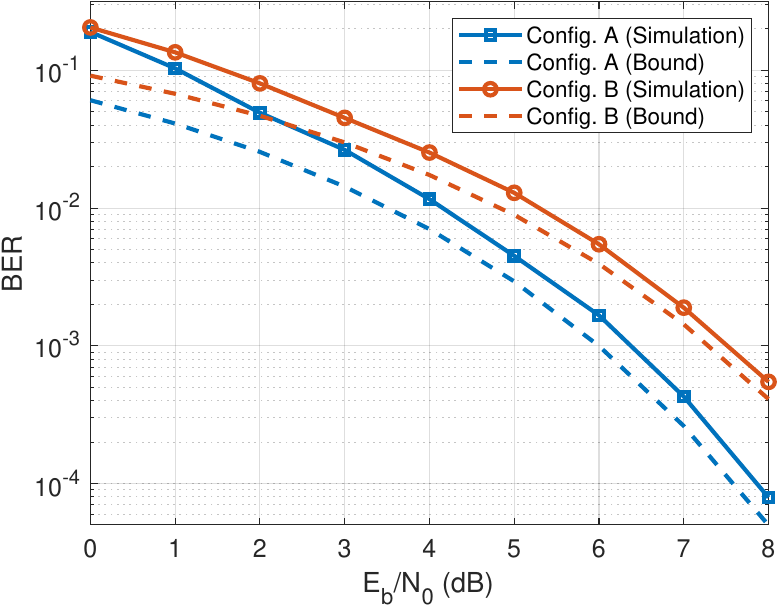}
    \caption{BER performance of Config. A and Config. B for $M=2$ in AWGN channel.}
    \label{fig:BER_M2}
\end{figure}

\subsection{SSB-FSK BER}
\label{subsecBER}
The authors in \cite{aulin1981continuous}, \cite{1094918}, showed that the minimum Euclidean distance is sufficient for illustrating the performance in terms of bit error probability for a large variety of CPM schemes. Since SSB-FSK is a new CPM scheme, in this section, we present the BER simulation to verify our derivations' effectiveness and tightness. We obtained the BER using the Viterbi algorithm, which is the optimal detector used to achieve MLSD for CPM schemes.\\
Fig. \ref{fig:BER_M2} shows the BER for Config. A and Config. B with $M=2$. For the sake of comparison, the union bound of each configuration is also given as a dashed line. It is shown that BER curves are approaching the bounds with high SNR values, which is consistent with what we previously stated, i.e., the union bound depicts the performance of the error probability for a high SNR regime. For low SNR, the bounds are quite loose. From Fig. \ref{fig:BER_M2}, we can conclude that the minimum Euclidean distance $d_{min}^2$ may be the right metric to adopt in order to depict the bit error probability performance.

\subsection{SSB-FSK CPM Power Spectrum Performance}
\label{subsec2}
Similar to Section~\ref{subsec1}, the effects of the parameters are already detailed for a wide variety of pulses in~\cite[Ch.\ 4]{anderson2013digital}. Therefore, we present only the effect of the parameter $w$ on the PSD of the SSB-FSK signals. Moreover, in this section, we also present the effect of $h$ on the SSB-LOSS.
 
\begin{figure}[t]
    \centering
        \includegraphics[width=\columnwidth]{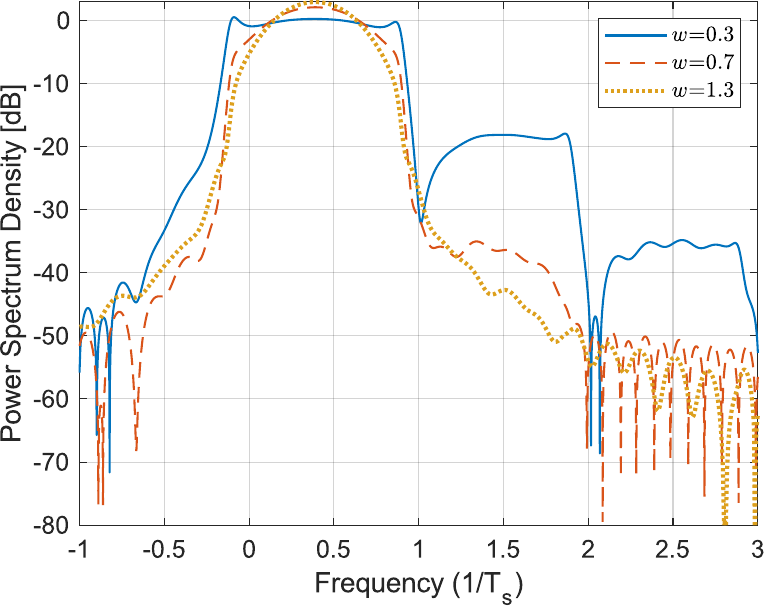}
        \caption{power spectral density of binary 5SSB-FSK with $h=0.78$ and for $w=0.3, 0.7, 1.3$}
        \label{fig:PSD_diffw}
    \end{figure}

Fig.~\ref{fig:PSD_diffw} illustrates the PSD of binary 5SSB-FSK for different pulse width $w=0.3, 0.7, 1.3$ as a function of the frequency, with $h=0.78$. The choices of the parameters ($L=5$ and $w=1.3$) are based on the first row of Table~\ref{Tab:opt_99} (Config.~A in Fig.~\ref{fig:Pareto optimum M2_999}), used also in Figs~\ref{fig:w_db} and~\ref{fig:h_dmin_w103} to depict the effect of the pulse width $w$ on the $d_{\text{min}}^2$.\\ 
In this part, we show the effect of pulse width $w$ on the power spectral density PSD. To clarify the selection of pulse width $w=1.3$ in our optimization solution, we added to Fig.~\ref{fig:PSD_diffw} two more plots for different pulse widths $w=0.3, 0.7$. The effect of the pulse width $w$ on the PSD is summarized in these three points:
\begin{itemize}
    \item it is clear that increasing $w$ accentuates the power exponential decay of the SSB-FSK modulated signal~\cite{fares2017power}.
    \item increasing the pulse width $w$ narrows the power spectral density, particularly in the frequency interval of width $1/T_{\text{s}}$, which is mostly clear when we compare the plots of pulse width $w=0.3$ and the others.
    \item Based on the closeup figure made around $h=0.78$ in Fig.~\ref{fig:w_db}, we can see that increasing $w$, in this particular region, is not decreasing $d_{\text{B}}^2$ that much (consequently $d_{\text{min}}^2$).
\end{itemize}
From these observations, it is clear that increasing the pulse width $w$ reduces the bandwidth $BW$ occupancy, improve the adjacent-channel interference, and maintains approximately the same $d_{\text{min}}^2$, based on the modulation index $h$ selected.\\
In general, we cannot increase the pulse width $w$ indefinitely. For instance, from Fig.~\ref{fig:w_db}, we can see that increasing the pulse width $w$ reduces the  $d_{\text{min}}^2$ especially for low modulation index $h$. Accordingly, the choice of parameters is a trade off aiming not only to optimize the $d_{\text{min}}^2$, but also the bandwidth $BW$ occupancy.

% Based on the closeup figure made around $h=0.78$ in Fig.~\ref{fig:w_db}, we can see that increasing $w$, in this particular region, is not decreasing $d_{\text{B}}^2$ that much (consequently $d_{\text{min}}^2$). Nevertheless, from Fig.~\ref{fig:PSD_diffw}, it is clear that increasing $w$ accentuates the power exponential decay of the SSB-FSK modulated signal~\cite{fares2017power}. Moreover, increasing the pulse width $w$ condenses the power spectral density, particularly in the frequency interval of width $1/T_{\text{s}}$, which is mostly clear when we compare the plots of pulse width $w=0.3$ and the others. Based on these observations, it is clear that increasing the pulse width $w$ will reduce the bandwidth $BW$ occupancy, and improve the adjacent-channel interference. However, we cannot increase the pulse width $w$ indefinitely. For instance, from Fig.~\ref{fig:w_db}, we can see that increasing the pulse width $w$ reduces the  $d_{\text{min}}^2$ specially for low modulation index $h$. Accordingly, we can understand the optimization choice; which is a practical tradeoff aiming not only to optimize the $d_{\text{min}}^2$, but also the bandwidth $BW$ occupancy. 

\begin{figure}[t]
    \centering
    \includegraphics[width=\columnwidth]{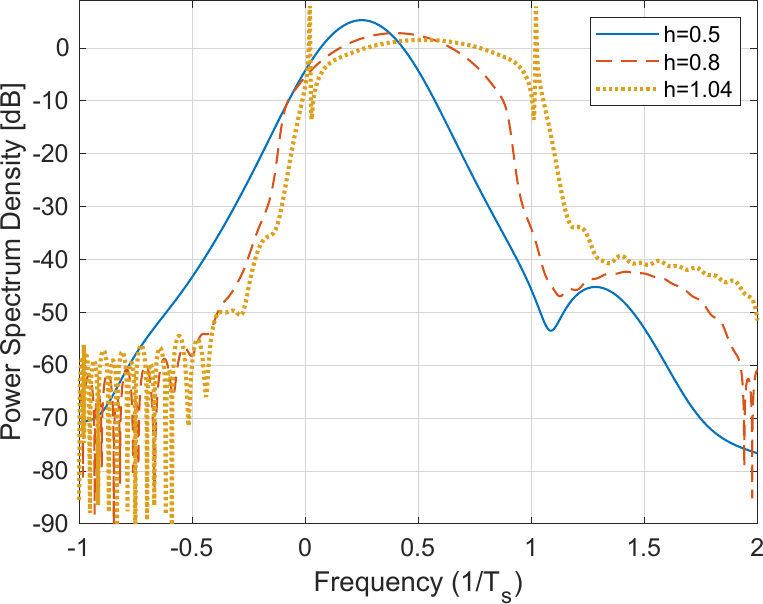}
    \caption{power spectral density of binary 12SSB-FSK with $w=0.8$ and $h=0.5, 0.8, 1.04$.}
    \label{fig:PSD_diffh}
\end{figure}

Fig.~\ref{fig:PSD_diffh} represents the PSD of the binary 12SSB-FSK, for $w=0.8$ and different $h=0.5, 0.8$ and $1.04$. In this part, we show the effect of the modulation index $h$ on the SSB-LOSS. As in the previous part, we used the parameters obtained from the optimization section (\ref{sec5}), i.e, the first row from Table \ref{Tab:opt_999} which corresponds to Config.~A$'$ in Fig.~\ref{fig:Pareto optimum M2_999} (result of the \emph{Pareto optimum} for $B_{999}$). In this particular context, we notice that we obtain the lowest SSB-LOSS. It is evident that increasing the modulation index $h$ increases the bandwidth $BW$ occupancy. It is especially visible for the frequency region between $0$ and $1/T_{\text{s}}$. On the other hand, the modulation index $h$ has the inverse effect on the SSB-LOSS: for $h=0.5$ we obtain an SSB-LOSS of $2.06\%$, and for $h=0.8$, the SSB-LOSS decreases to $1.63\%$, and then for $h=1.04$ it reached the lowest value with an SSB-LOSS of $0.366\%$. However, for different parameters, e.g., 6SSB-FSK and pulse width $w=0.37$, increasing the modulation index will increase the SSB-LOSS. Therefore, it is impossible to give a general trend about the modulation index $h$ on the SSB-LOSS. Otherwise, it is always true that taking an integer modulation index $h$ will reduce the SSB-LOSS; this decrease is due to the spectral lines', which take a part of the transmitted power. This observation is due to selecting a modulation index $h$ around an integer ($\approx 1$). Even if these spikes seem undesirable because a part of the transmitted power is wasted, they could be of great interest for synchronization purposes \ref{sec:advantage_synchro}~\cite[Ch.\ 9, Sec.\ 1]{anderson2013digital}\cite{Xu2019}.

\subsection{Single-Sideband CPM Energy-Bandwidth Comparison}
\label{subsec3}
In this section, we show comparison plots between SSB-FSK and other CPM schemes. We used SSB-FSK parameters based on the results obtained from the optimization study given in Section~\ref{sec5}. The plots are characterized by the SSB bandwidth occupancy in the x-axis ($BW$), and $10\log_{10}(d_{\text{min}}^{2}/2)$ for the y-axis (energy). Moreover, we divided this section into two parts; one illustrates the occupied $BW$ as $99\%$ of the power inside $BT_{\text{b}}$. The other illustrates $99.9\%$ of the power in the $BT_{\text{b}}$ (Note the $99.9\%$ is a larger measure compared to the $99\%$; therefore, $99.9\%$ curves and configurations are located on the right side compared to $99\%$). The curves and configurations with better performance are located toward the upper left side. For each $BW$ parts, we will always start with a general comparison between the optimum curves obtained for SSB-FSK with the other CPM schemes, where the optimum curves do not take into consideration the complexity ($N_{\text{s}}$). 
Then we move on to a more detailed comparison, where we also consider the configurations with optimized complexity and how much they are biased from the optimum curves. The optimum curves for SSB-FSK are the same \textit{Pareto optimum} curves obtained from Section.\ref{sec:opt_1}, and they are presented with the same notation $P_{\text{opt}-\text{max}}$, followed by the modulation level $M$ (e.g., $P_{\text{opt}-\text{max}-4}$, for $M=4$). For each of the optimum curves, the optimum normalized minimum squared Euclidean distance $d_{\text{min}-\text{opt}}^2$ and the optimum bandwidth $BW$ occupancy $BW_{\text{opt}}$ are also presented in the plots, denoted by Config.~A and Config.~B for $B_{99}$ respectively. Likewise, they will be denoted by Config.~A$'$ and Config.~B$'$ for $B_{999}$. A subscription is used to refer to the modulation level, e.g., Config.~$\text{A}_2$ refers to Config.~A for the binary case. Moving on to SSB-FSK with optimized $N_{\text{s}}$, we select the same configurations obtained from Section.\ref{sec:opt_2}.

\begin{figure}[t]
    \centering
        \includegraphics[width=\columnwidth]{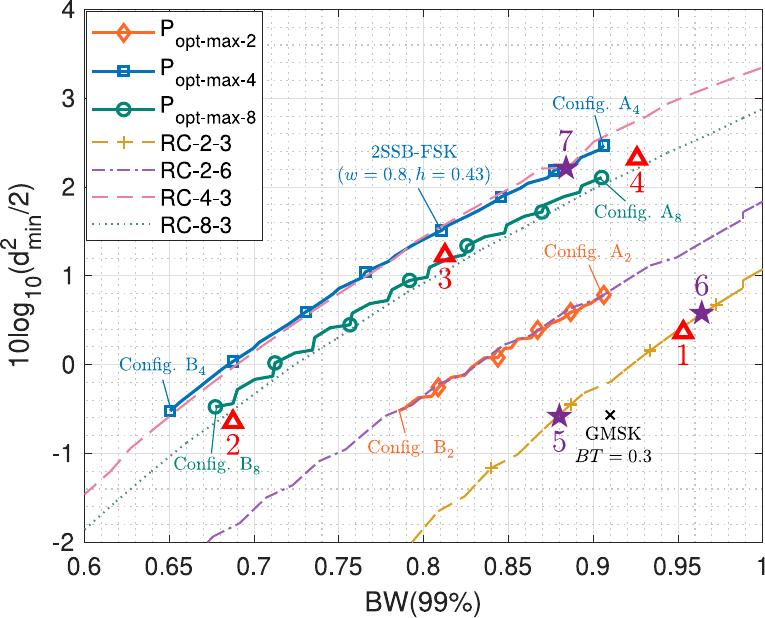}
        \caption{Energy-Bandwidth plot for $L$SSB-FSK and $L$RC for different modulation level $M=2,4,8$ using $B_{99}$. GMSK is presented as an``x" mark.}
        \label{fig:B99}
    \end{figure}
\begin{table}[]
    \centering
        \resizebox{0.9\columnwidth}{!}{%
\begin{tabular}{ccccc}
    \toprule
Index & \begin{tabular}[c]{@{}c@{}}Modulation\\ Type\end{tabular}                   & $d_{min}^2$ & $BW$       & States ($N_{\text{s}}$) \\ \midrule \midrule
$1$   & \begin{tabular}[c]{@{}c@{}}1SSB-FSK\\ ($M=8$,$h=0.25$,$w=0.7$)\end{tabular} & $2.175$     & $0.953$  & $4$                     \\ \hline
$2$   & \begin{tabular}[c]{@{}c@{}}2SSB-FSK\\ ($M=8$,$h=0.25$,$w=0.5$)\end{tabular} & $1.724$     & $0.687$  & $32$                    \\ \hline
$3$   & \begin{tabular}[c]{@{}c@{}}2SSB-FSK\\ ($M=4$,$h=0.4$,$w=0.5$)\end{tabular}  & $2.654$     & $0.8125$ & $20$                    \\ \hline
$4$   & \begin{tabular}[c]{@{}c@{}}2SSB-FSK\\ ($M=4$,$h=0.5$,$w=1$)\end{tabular}    & $3.412$     & $0.925$  & $8$                     \\ \hline
$5$   & \begin{tabular}[c]{@{}c@{}}3RC\\ ($M=2$,$h=0.5$)\end{tabular}               & $1.75$      & $0.88$   & $16$                    \\ \hline
$6$   & \begin{tabular}[c]{@{}c@{}}3RC\\ ($M=2$,$h=0.6$)\end{tabular}               & $2.286$     & $0.964$  & $40$                    \\ \hline
$7$   & \begin{tabular}[c]{@{}c@{}}3RC\\ ($M=4$,$h=0.5$)\end{tabular}               & $3.33$      & $0.884$  & $64$                    \\ \bottomrule
\end{tabular}
        }
\captionsetup{justification=centering}
            \caption{\\ Performance of $L$SSB-FSK and $L$RC configurations with the lowest complexity for $99\%$ $BW$ occupancy.}
            \label{Tab:New_conf_99}
\end{table}

\begin{figure}[t]
            \centering
                \includegraphics[width=1.03\columnwidth,]{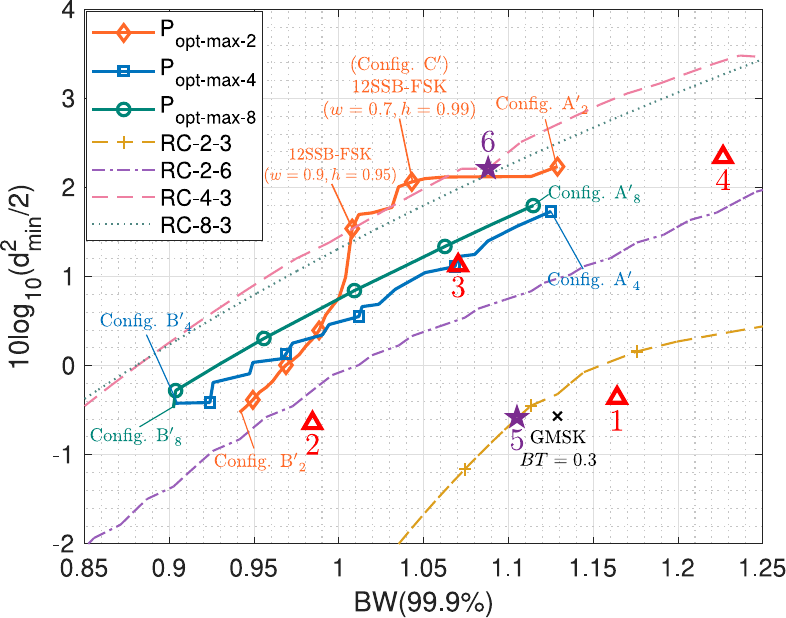}
                \caption{Energy-Bandwidth plot for $L$SSB-FSK and $L$RC for different modulation level $M=2,4,8$ using $B_{999}$. GMSK is presented as an ``x" mark.}
                \label{fig:B999}
        \end{figure}

These configurations are presented in Fig. \ref{fig:B99} and Fig. \ref{fig:B999} as triangular marker with an index number. Similar to Section\ref{sec:opt_2}, the index number presents the number of the row in Table.\ref{Tab:New_conf_99} and Table.\ref{Tab:New_conf_999} for $99\%$ and $99.9\%$ $BW$ occupancy respectively. For a fair comparison with other CPM schemes, we present the RC curves for different modulation levels $M$ and pulse lengths $L$. The RC pulse's legends are given starting with the pulse type followed by the modulation level $M$, and the pulse length $L$ (e.g., RC-2-3 is the raised cosine pulse for modulation level $M=2$ and pulse length $L=3$). For each RC curves, we also present the configurations with $N_{\text{s}}<64$ (similar to Section.\ref{sec:opt_2}, configurations with $N_{\text{s}}>64$ are considered as high complex system). 

        \begin{table}[]
            \centering
            \resizebox{0.9\columnwidth}{!}{%
            \begin{tabular}{ccccc}
            \toprule 
            Index & \begin{tabular}[c]{@{}c@{}}Modulation\\ Type\end{tabular}                   & $d_{min}^2$ & $BW$      & States ($N_{\text{s}}$) \\ \midrule \midrule
            $1$   & \begin{tabular}[c]{@{}c@{}}2SSB-FSK\\ ($M=2$,$h=0.5$,$w=0.6$)\end{tabular}  & $1.84$      & $1.164$ & $4$                     \\ \hline
            $2$   & \begin{tabular}[c]{@{}c@{}}2SSB-FSK\\ ($M=8$,$h=0.25$,$w=0.5$)\end{tabular} & $1.724$     & $0.984$ & $32$                    \\ \hline
            $3$   & \begin{tabular}[c]{@{}c@{}}2SSB-FSK\\ ($M=4$,$h=0.4$,$w=0.6$)\end{tabular}  & $2.6$       & $1.07$  & $20$                    \\ \hline
            $4$   & \begin{tabular}[c]{@{}c@{}}6SSB-FSK\\ ($M=2$,$h=1$,$w=1.1$)\end{tabular}    & $3.43$      & $1.226$ & $32$                    \\ \hline
            $5$   & \begin{tabular}[c]{@{}c@{}}3RC\\ ($M=2$,$h=0.5$)\end{tabular}               & $1.75$      & $1.105$ & $16$                    \\ \hline
            $6$   & \begin{tabular}[c]{@{}c@{}}3RC\\ ($M=4$,$h=0.5$)\end{tabular}               & $3.33$      & $1.088$  & $64$                    \\ \bottomrule
            \end{tabular}
            }
            \captionsetup{justification=centering}
            \caption{\\ Performance of $L$SSB-FSK and $L$RC configurations with the lowest complexity for $99.9\%$ $BW$ occupancy.}
            \label{Tab:New_conf_999}
            \end{table}
            
These configurations are shown as star marker with an index number, where similarly to SSB-FSK the index number presents the number of the row in Table.\ref{Tab:New_conf_99} and Table.\ref{Tab:New_conf_999} respectively for $99\%$ and $99.9\%$ $BW$ occupancy. Finally, the configuration of GMSK with $BT=0.3$ is also shown as ``X" marker in Fig. \ref{fig:B99} and Fig. \ref{fig:B999}. The GMSK has a complexity $N_{\text{s}}=16$. Note, all configurations located outside the region of the energy-bandwidth comparison are not considered ($y$-axis $<-2$ dB for $99\%$ and $99.9\%$, the $x$-axis $>1$ and $>1.3$ for $99\%$ and $99.9\%$ respectively).
\subsubsection{99\% Bandwidth occupancy}
Fig.~\ref{fig:B99} illustrates the performance for $B_{99}$ of SSB-FSK signals compared to $L$RC for different modulation levels M. Besides, the GMSK signal for $BT=0.3$ is also shown as x mark in the plots. Starting with the SSB-FSK curves, the $P_{\text{opt}-\text{max}}$ for all modulation levels $M$ outperforms the GMSK, where we obtain an energy consumption gain of $1.35$, $3.03$, and $2.67$ dB, for respectively Config.~$\text{A}_2$, Config.~$\text{A}_4$, and Config.~$\text{A}_8$ for approximately the same bandwidth $BW$ occupancy. 
Similarly, we obtain a bandwidth $BW$ occupancy gain of $0.125$, $0.26$, and $0.233$, respectively for Config.~$\text{B}_2$, Config.~$\text{B}_4$, and Config.~$\text{B}_8$ for almost the same energy consumption. Moving to the binary RC, the $P_{\text{opt}-\text{max}-2}$ outperforms the 2RC; however, we obtain the same performance compared to binary 6RC. On the other hand, the $P_{\text{opt}-\text{max}-4}$ and $P_{\text{opt}-\text{max}-8}$ outrun the binary RC for all cases. Likewise, for $M=4$, using the $P_{\text{opt}-\text{max}-4}$, we obtain almost the same performance as quaternary 3RC. Similarly, for $M=8$, we obtain a slightly better performance using $P_{\text{opt}-\text{max}-8}$ compared to $8$-ary 3RC. However, for all bandwidth occupancies, the $P_{\text{opt}-\text{max}-4}$ outperforms the $P_{\text{opt}-\text{max}-8}$, meaning that increasing the modulation level $M$ will not always increase the performance. Besides, we can highlight the effect of increasing the modulation level $M$ especially between $2$ and $4$; we obtain an energy consumption gain of $1.68$ dB between Config.~$\text{A}_2$ and Config.~$\text{A}_4$ for the same bandwidth occupancy. Similarly, considering now the bandwidth occupancy, we obtain a gain of $0.135$ between Config.~$\text{B}_2$ and Config.~$\text{B}_4$ for almost the same energy consumption.
Moving on to SSB-FSK configurations with optimized complexity. As observed from Fig. \ref{fig:B99}, configuration ``$5$" and GMSK have similar energy performance and the same number of states $N_{\text{s}}=16$, while ``$5$" has a slightly better $BW$ occupancy with a difference of $\approx 0.03$, which makes ``$5$" the best choice. Considering now the configuration ``$1$", we obtain an energy gain of $\approx 0.93$ dB with a slight increase in $BW$ occupancy of $\approx 0.043$ and $\approx 0.073$ compared to GMSK and ``$5$" respectively. In complexity, ``$1$" exhibits the lowest number of state $N_{\text{s}}=4$ compared to $N_{\text{s}}=16$. Configuration ``$6$" shows a similar performance to ``$1$'', while $1$ offers a remarkably lower complexity in comparison to ``$6$" having $N_{\text{s}}=40$,  which makes ``$1$" the best choice compared to ``$5$",``$6$" and the GMSK. In the low $BW$ occupancy area, we only have configuration ``$2$" with a similar performance to Config.~$\text{B}_8$ and with a moderate complexity $N_{\text{s}}=32$. By comparing configuration ``$3$"  with $2$SSB-FSK ($M=2,w=0.8$ and $h=0.43$), we can note that both configurations have similar $BW$ occupancy, while ``$3$" has slightly lower energy performance ($\approx 0.3$) dB. Nevertheless, configuration ``$3$" has extremely lower complexity ($N_{\text{s}}=20$) in comparison to ($N_{\text{s}}=200$). Finally, we compare the configurations ``$7$" and ``$4$". Both configurations have similar energy performance, while ``$4$" has slightly worst $BW$ occupancy with an increase of $\approx 0.04$. However, in terms of complexity, ``$4$" has a notably lower number of states ($N_{\text{s}}=8$) compared to ``$7$" ($N_{\text{s}}=64$). Moreover, ``$4$" reveals a similar $BW$ occupancy to GMSK, while ``$7$" has a considerable energy gain of $\approx 2.9$ dB and two times lower complexity in comparison with GMSK.
Overall, these comparisons reveal that the configurations obtained for the SSB-FSK suggest the best choice in terms of energy-bandwidth and receiver complexity, especially configuration ``$4$".

\subsubsection{99.9\% Bandwidth occupancy}
\label{sec:Comp_999}
As in the previous part, in Fig. \ref{fig:B999}, we compare SSB-FSK $P_{\text{opt}-\text{max}}$ curves with GMSK (shown as x mark in the plot). We achieve a gain of $2.8$ dB, $2.3$ dB, and $2.4$ dB for Config.~$\text{A}'_2$, Config.~$\text{A}'_4$, and Config.~$\text{A}'_8$ respectively at nearly the same bandwidth $BW$ occupancy. Likewise, we obtain a gain of $0.19$, $0.227$, and $0.226$, for Config.~$\text{B}'_2$, Config.~$\text{B}'_4$, and Config.~$\text{B}'_8$ respectively at almost the same energy consumption. Config.~C$'$ is reported for sake of comparison and it is clearly an interesting operating point as discussed in the optimization study given in Section~\ref{sec5}. Besides, the $P_{\text{opt}-\text{max}}$ for different modulation levels $M$ outperform the binary RC for pulse length $L=3$ and $6$. However, the pulse length $L$ used in $P_{\text{opt}-\text{max}-2}$ is always larger than or equal to $6$, specially for Config.~$\text{A}'_2$ where we have a pulse length $L=12$. With $L=12$, we have a disastrous impact on the receiver complexity. The solution to treat the complexity problem is to use some other configurations with lower complexity $N_{\text{s}}$ at the cost of some performance drop.
For modulation level $M=4$, we note a different behavior compared to the previous part. In $B_{99}$, we obtained the same performance using $P_{\text{opt}-\text{max}-4}$ and quaternary 3RC. However, for $B_{999}$, the quaternary 3RC outruns the $P_{\text{opt}-\text{max}-4}$ and $P_{\text{opt}-\text{max}-8}$ for all configurations. Unlike what we noticed for the quaternary case, we do not have a unique behavior for the binary case. For instance, the $P_{\text{opt}-\text{max}-2}$ outperforms the $P_{\text{opt}-\text{max}-4}$ for all $BT_{\text{b}}\geq 1$, and outperforms the $P_{\text{opt}-\text{max}-8}$ for all $BT_{\text{b}}\geq 1.01$. Moreover, it also has a similar or better performance than the quaternary 3RC for points defined on the $B_{999}$ plot.
For SSB-FSK configurations with optimized complexity, the configuration ``$1$", ``$5$" and the GMSK have similar energy performance and slightly different $BW$, with ``$1$" having the worst $BW$ between the configurations. Besides, configuration ``$1$" has a lower receiver complexity ($N_{\text{s}}=4$) in comparison to the two other configurations ($N_{\text{s}}=16$), which makes ``$1$" the best choice among these latest three ( configurations ``1" and ``5" and GMSK). Similarly, to the previous study, in the low $BW$ region, we have configuration ``$2$" with similar energy to Config.~$\text{B'}_2$ and with slightly higher $BW$ occupancy of $0.043$ (which remains very small). Moreover, Config.~``$2$" has a moderate complexity of $N_{\text{s}}=32$. Now, we compare Configurations ``$3$" and ``$6$", where both exhibits similar $BW$ occupancy, while ``$6$" shows higher energy performance with a gain of $\approx 1.08$ dB compared to ``$3$". This is offered at the expense of complexity, indeed, ``$3$" offers a remarkably lower complexity ($N_{\text{s}}=20$) in comparison to ``$6$" ($N_{\text{s}}=64$). Finally, we compare the configurations ``$6$" and ``$4$". Both configurations have similar energy performance. On the other hand, ``$4$" shows higher $BW$ occupancy compared to ``$6$" with a difference of $\approx 0.136$. In terms of complexity, ``$4$" has the lowest complexity ($N_{\text{s}}=32$) in comparison to ``$6$" ($N_{\text{s}}=64$), where ``$4$" needs only half the number of states.

Overall, if we consider only the receiver complexity, the SSB-FSK with configuration ``$1$" presents the best choice. Nevertheless, in terms of energy-bandwidth and receiver complexity, it is hard to consider only one configuration since ``$3$", ``$4$", and ``$6$" show different tradeoffs between complexity and performance. Therefore, these three are considered the best choices for $99.9\%$ $BW$ occupancy.

Table ~\ref{Tab:Compare} presents a performance comparison between  6SSB-FSK ($h=1$ and $w=1.1$), which is configuration ``$4$" from Table.~\ref{Tab:New_conf_999}; and 12SSB-FSK ($h=1$, $w=0.37$), which is the study case from the original proposal in~\cite{fares2018quantum,fares2017new}.  Based on Table~\ref{Tab:Compare},  it is clear that  configuration ``$4$" outperforms the original proposal. More precisely, ``$4$" offers a better spectrum occupancy, whereas maintaining a better energy consumption performance ($\approx 2.56$ dB ). On the other hand, the original proposal present a slightly better SSB-LOSS with a difference of $\approx 0.3$.

\begin{table}[]
    \centering
    \resizebox{0.9\columnwidth}{!}{%
    \begin{tabular}{ccccc}
        \toprule
    Modulation Type                                                               & $d_{min}^2$ & $BW$ ($99.9\%$) & SSB-LOSS & $N_{\text{s}}$ \\ \midrule \midrule 
    \begin{tabular}[c]{@{}c@{}}12SSB-FSK (original proposal)\\ ($M=2$, $h=1$, $w=0.37$)\end{tabular} & $1.9$       & $2.06$          & $0.469$  & $2048$         \\ \hline
    \begin{tabular}[c]{@{}c@{}}6SSB-FSK (configuration ``$4$")\\ ($M=2$,$h=1$,$w=1.1$)\end{tabular}   & $3.43$      & $1.226$         & $0.77$   & $32$           \\ \bottomrule
    \end{tabular}
    }
    \captionsetup{justification=centering}
            \caption{\\Performance comparison with the original SSB-FSK proposal.}
            \label{Tab:Compare}
    \end{table}
    \subsection{Integer Modulation Index Synchronization Advantage}
    \label{sec:advantage_synchro}
\begin{figure}[t]
    \centering
        \includegraphics[width=1.03\columnwidth,]{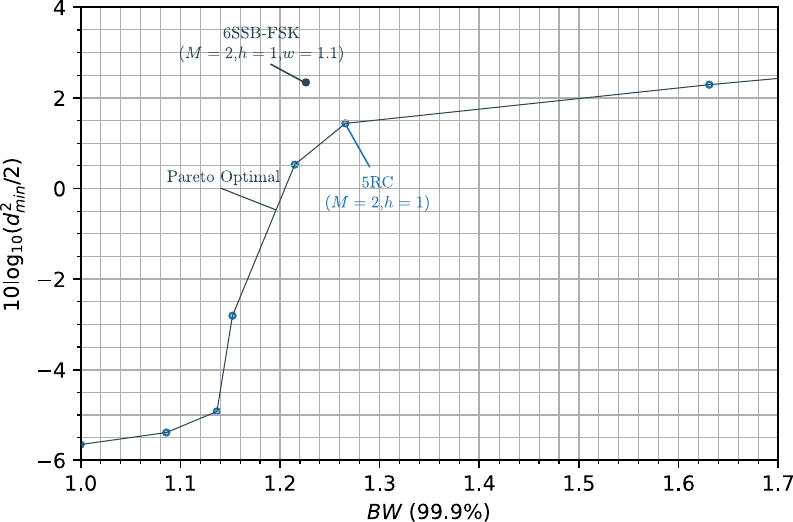}
        \caption{ \textit{Pareto optimum} plot of RC with modulation index $h=1$ in comparison to configuration ``$4$" for $B_{999}$.}
        \label{fig:Popt_RC}
\end{figure}

    In~\cite[Ch.\ 9, Sec.\ 1]{anderson2013digital} and \cite{Xu2019}, the authors show an advantage of using integer modulation index $h$ for synchronization. Based, on the results obtained from Section.\ref{sec:Comp_999}, configuration ``$4$" has an integer modulation index $h=1$ and it is one of the best configurations obtained for $B_{999}$. Usually, CPM schemes with integer $h$ are avoided due to there weak performance \cite{anderson2013digital}, which is not the case for configuration ``$4$". \\
    Since we present a comparison between SSB-FSK and RC in Section.\ref{sec6}, it is interesting to show how the RC scheme behave when it is operating with an integer $h$. In Fig.\ref{fig:Popt_RC}, we present the \textit{Pareto optimum} plot of RC with modulation index $h=1$
    for all pulse length $1 \le L \le 12$ and modulation level $M =\{2,4,8\}$. Integer modulation indices $h>1$ are not considered, since they show a poor $BW$ occupancy. From Fig.\ref{fig:Popt_RC} we can observe that configuration ``$4$" presents a gain of $0.9$ dB and slightly lower $BW$ occupancy $0.04$ in comparison to the best configuration presented in the \textit{Pareto optimum} for RC ($L=5$,$M=4$,$h=1$). In terms of complexity, the two configurations have the same number of state $N_{\text{s}}= 32$. Note, all configurations shown on the \textit{Pareto optimum} curve (other than the selected configuration of RC) has a number of states $N_{s}>128$ except the first configuration located at the lower-left corner has a $N_{s}=2$ (at the expense of mediocre performance).
    
\section{Concluding Remarks and Design Directives}
\label{sec+}
From the optimization methods, we have selected the best parameter combinations to obtain the maximum performance of the SSB-FSK scheme. Based on the parameters chosen from the optimization solution, we achieved a better spectrum occupancy, a gain in error probability performance, and a massive decrease in complexity, compared to the original SSB-FSK proposal~\cite{fares2018quantum,fares2017new}. Our optimization method leads to practical tradeoffs through certain parameter combinations. These tradeoffs differ because they prioritize performance metrics differently. The choice depends on the target application and, consequently, the resources we want to allocate. Finally, we have applied an energy-bandwidth comparison, where we achieved a similar or better performance compared to certain robust CPM modulations. In general, the SSB-FSK has the advantage of being more flexible than RC. For instance, it has been proven that this new waveform is highly tunable, able to provide the most suitable configuration responding to the designer requirement (energy, bandwidth, or complexity). In $B_{99}$ case, we showed that the configurations obtained from SSB-FSK modulation outperform the RC and GMSK in energy-bandwidth and receiver complexity in any position in the defined comparison region. It is always possible to find a SSB-FSK configuration that outperforms the RC. For $B_{999}$, the answer cannot be binary or clear-cut, but a more nuanced analysis has been made, offering several possible tradeoffs.
Nevertheless, the B999 analysis allows us to define one SSB-FSK configuration combining excellent performance and advantage in synchronization since it operated with an integer modulation index. This, in particular, offers to SSB-FSK scheme the potential to go beyond what has been achieved with well known CPM schemes (e.g., RC and GMSK). Moreover, we showed that using an integer $h$ with RC will never exceed the SSB-FSK results. Overall, it is clear that the SSB-FSK is not always the best scheme, and it heavily depends on the parameters. Therefore, the choice to select the best CPM schemes heavily rely on the target application.

\section{Conclusion}
\label{sec7}
In this paper, we have investigated a CPM modulation that can directly generate a SSB spectrum. To exploit this modulation scheme's characteristics, we investigated the error probability based on the derivation of the minimum Euclidean distance as a function of the modulation index, pulse width, and pulse length. We explored the spectrum of SSB-FSK by quantifying its signal power bandwidth occupancy using a numerical method. 
We addressed the receiver complexity aspects. All these metrics have been used in two different optimization methods to illustrate this waveform's full potential alongside well-known CPM schemes. Future work will focus on the sub-optimum Viterbi-based demodulation scheme using pulse amplitude modulation (PAM) decomposition to offer more sizable complexity reduction. The aim will be to explore other possible configurations offering further potential gains to go beyond what has been achieved with well known CPM schemes.

% SSB-FSK scheme with modulation level $M=4$ gives the best performance compared with other SSB-FSK modulation levels $ M $ as it was expected in~\cite{aulin1981continuous}. 
% For $B_{999}$ case, the binary 12SSB-FSK presents an excellent performance: the advantage of this particular scheme is twofold. From one side, this scheme is the only modulation with good performance (similar to RC) while having almost an integer modulation index $h\approx1$ (in the $B999$ case): with integer modulation indices, other pulses behave as weak modulation schemes \cite{anderson2013digital}. From another side, using approximately integer modulation indices generates the spectral lines (\textit{spikes}), which can be efficiently utilized in synchronization in comparison with other modulations (e.g., RC) \cite{Xu2019}. Therefore, combining these two aspects, we think that SSB-FSK can go beyond what has been achieved with well-known CPM schemes when considering all performance metrics.

\appendices
\section{}
\label{FirstAppendix}
In this appendix, we detail how (\ref{dmin}) was obtained from (\ref{eq_dmin_1}), similarly to the derivation given in \cite[ch.2, p.26-p.28]{anderson2013digital} and \cite[ch.4, p.251-p.252]{proakis2008digital}.\\
The carrier modulated signal corresponding to the real part of (\ref{eq:Modulation}) is expressed as
\begin{align}
    \label{eq:Modualtion_Carrier}
    s(t,\alpha) = \sqrt{\frac{2E}{T_{s}}} \text{cos}[w_{0}t + \phi(t;\alpha)]
\end{align}
where $E_{\text{s}}= 2E$ is the energy per transmitted symbol and $w_{0}$ is the carrier frequency. To simplify the derivation, the carrier frequency $w_{0}$ is assumed much larger than $2\pi/T_{\text{s}}$.
We suppose that the two signals $s(t,\alpha_{i})$ and $s(t,\alpha_{j})$ differ over an interval $N$. The Euclidean distance between the two signals over the interval $N$ is defined as \cite{proakis2008digital}
\begin{align}
    \begin{aligned}
        \label{eq:Euclidean_define}
    &\int_{0}^{NT_{\text{s}}}[s(t,\alpha_{i})-s(t,\alpha_{j})]^2dt =\\
    &\int_{0}^{NT_{\text{s}}} s(t,\alpha_{i})^2 dt + \int_{0}^{NT_{\text{s}}} s(t,\alpha_{j})^2dt \\
    &-2 \int_{0}^{NT_{\text{s}}}  s(t,\alpha_{i})s(t,\alpha_{j})\;dt.
    \end{aligned} 
\end{align}
Since both signals are constant envelope with energy $E$ over an interval $T_{\text{s}}$ (we obtain $E$ by integrating the square of (\ref{eq:Modualtion_Carrier}) over an interval $T_{\text{s}}$), the second line of (\ref{eq:Euclidean_define}) can be expressed as
\begin{align}
    \label{eq:part1}
     \int_{0}^{NT_{\text{s}}} s(t,\alpha_{i})^2 dt + \int_{0}^{NT_{\text{s}}} s(t,\alpha_{j})^2dt = 2 N E.
\end{align}
Using the trigonometric function $\text{cos}(a)\text{cos}(b)=\frac{1}{2}(\text{cos}(a-b)+cos(a+b))$ and assuming large carrier frequency, the third line in (\ref{eq:Euclidean_define}) can be presented as 
\begin{align}
    \begin{aligned}
        \label{eq:part2}
        &2 \int_{0}^{NT_{\text{s}}}  s(t,\alpha_{i})s(t,\alpha_{j})\;dt =\\
        &2\frac{2E}{T_{s}}\int_{0}^{NT_{\text{s}}}\text{cos}[w_{0}t + \phi(t;\alpha_{i})]\text{cos}[w_{0}t + \phi(t;\alpha_{j})]=\\
        &\frac{2E}{T_{\text{s}}} \int_{0}^{NT_{\text{s}}} \text{cos}[\phi(t;\alpha_{i})-\phi(t;\alpha_{j})]dt.
    \end{aligned}
\end{align}
From (\ref{eq:part1}) and (\ref{eq:part2}), we can present (\ref{eq:Euclidean_define}) in the form
\begin{align}
    \label{eq:Euclidean_part1}
    2NE-\frac{2E}{T_{\text{s}}} \int_{0}^{NT_{\text{s}}} \text{cos}\Delta\phi(t)dt,
\end{align}
where $\Delta\phi(t)$ denotes the phase difference between the two signals.
Here $E$ is the energy obtained for an interval and not for a data bit. Therefore, for a fair comparison between many modulation schemes with different modulation levels $M$, $E$ has to be normalized to obtain the energy per data bit $E_{b}$. The normalization of $E$ is defined as
\begin{align}
    \label{eq:Euclidean_part2}
    E=\text{log}_{2}ME_{\text{b}}.
\end{align}
Placing (\ref{eq:Euclidean_part1}) and (\ref{eq:Euclidean_part2}) in (\ref{eq_dmin_1}), we obtain (\ref{dmin}).
\balance
\bibliography{References}
\bibliographystyle{IEEEtran}
\balance
\end{document}